\documentclass[aps,prl,reprint,footinbib,longbibliography,floatfix]{revtex4-2}

\usepackage{amsmath,amssymb,mathtools,bm}
\usepackage{amsthm}
\usepackage{graphicx}
\usepackage{microtype}
\usepackage{hyperref}
\hypersetup{colorlinks=true,citecolor=blue,urlcolor=blue,linkcolor=blue}

\newcommand{\Tr}{\operatorname{Tr}}
\newcommand{\EF}{E_{\mathrm F}}
\newcommand{\GEF}{E_{\mathrm F}^{\mathrm G}}
\newcommand{\EC}{E_{\mathrm C}}
\newcommand{\TMSV}{\lvert\Psi_q\rangle}
\newcommand{\Dop}{\widehat D_t}
\newcommand{\cE}{\mathcal E}
\newcommand{\cF}{\mathcal F}

\usepackage{times,txfonts}

\begin{document}

\title{Optimality of Gaussian Entanglement of Formation}

\author{Gerardo Adesso}
\affiliation{School of Mathematical Sciences and Centre for the Mathematics and Theoretical Physics of Quantum Non-Equilibrium Systems, University of Nottingham, University Park, Nottingham NG7 2RD, United Kingdom}

\date{\today}

\begin{abstract}
We prove that the entanglement of formation of every two-mode Gaussian state coincides with its Gaussian restriction, solving a longstanding open problem in continuous variable quantum information theory. The key result is a sharp affine relation between entanglement and a generalized Einstein--Podolsky--Rosen observable, valid for arbitrary pure two-mode states, including non-Gaussian ones. We further prove that the entanglement of formation is additive on arbitrary tensor products of two-mode Gaussian states, and hence coincides with the entanglement cost, settling the latter for all two-mode Gaussian states. As a consequence, we obtain in closed form the entanglement cost of thermal-loss, thermal-amplifier and additive-noise bosonic channels. Our results, which also extend to bisymmetric multimode states, establish that no two-mode state is less entangled than the Gaussian state with the same covariance matrix, yielding measurable lower bounds on both the entanglement of formation and the entanglement cost of arbitrary non-Gaussian states.
\end{abstract}
\maketitle

\textit{Introduction.---}
Gaussian states are the workhorses of continuous variable quantum technologies, thanks to a combination of broad practical relevance with a frugal mathematical and experimental description: a two-mode Gaussian state is completely specified by its first moments and a $4\times4$ covariance matrix  \cite{Weedbrook2012,Adesso2014}. Despite their limitations \cite{leah2025}, Gaussian states underpin relevant applications including  teleportation, communication networks, distributed quantum sensing, integrated photonics, and hybrid microwave--optical interfaces \cite{Braunstein2005,Weedbrook2012,Adesso2014,Furusawa1998,Grosshans2003,Guo2020}, as demonstrated in recent experiments  \cite{Jia2025,Lou2024,Sahu2023,Meesala2024,Zhao2025}.

A central quantity to benchmark the resource power of quantum states  in this setting is the {\em entanglement of formation}  \cite{Bennett1996}. For a bipartite state $\rho$,
\begin{equation}
 \EF(\rho)=\inf_{\{p_j,\psi_j\}}\sum_j p_j E(\psi_j),
 \qquad
 \rho=\sum_jp_j\lvert\psi_j\rangle\!\langle\psi_j\rvert,
 \label{eq:eof}
\end{equation}
where $E(\psi_j)$ is the entropy of either reduced state of the pure vector $\lvert\psi_j\rangle$. Thus $\EF$ is the least average pure-state entanglement needed to prepare $\rho$ by mixing pure states; its regularization is the entanglement cost $\EC(\rho)$ \cite{Hayden2001}, the minimum asymptotic entanglement per copy needed to prepare a state by local operations and classical communication. For a Gaussian state $\rho_V$ with covariance matrix $V$, the {\em Gaussian entanglement of formation} $\GEF(V)$, also denoted $\GEF(\rho_V)$, restricts Eq.~\eqref{eq:eof} to ensembles of Gaussian pure states \cite{Wolf2004}. Since this is a restriction of the optimization, in general
\begin{equation}
 \EF(\rho_V)\leq\GEF(V),
 \label{eq:trivialordering}
\end{equation}
and the question is whether a non-Gaussian ensemble can make the inequality strict. Note that being Gaussian is a property of the density matrix, not of the ensembles realizing it, which are in general non-Gaussian. This question was listed in 2005 as \href{https://github.com/google-deepmind/formal-conjectures/issues/3446}{Open Quantum Problem 29} \cite{KruegerWerner2005}.

The equality $\EF(\rho_V)=\GEF(V)$ in (\ref{eq:trivialordering}) was proven for symmetric two-mode states, where a balanced Einstein--Podolsky--Rosen (EPR) observable and mode-exchange symmetry reduce the problem to a one-dimensional extremality statement \cite{Giedke2003}. Several general solutions were subsequently proposed \cite{Marian2008,IvanSimon2008,Akbari2015}, but the central question remained unresolved. Marian and Marian constructed the relevant Gaussian candidate decomposition \cite{Marian2008}, but its optimality over non-Gaussian ensembles was assumed rather than established. Ivan and Simon isolated the missing ingredient as an extremality conjecture for a generalized EPR observable with unequal gains \cite{IvanSimon2008}; a subsequent rearrangement argument proved it under an additional condition on the Schmidt coefficients \cite{Akbari2015}. Later analytical bounds and efficient numerical procedures made the Gaussian optimization tractable \cite{AdessoIlluminati2005,Tserkis2017,Wilde2018,Tserkis2019}, while its identification with the unrestricted convex roof remained unproven.

In this Letter, we remove this final obstruction.  We further prove that the entanglement of formation is \emph{additive} on arbitrary tensor products of two-mode Gaussian states, and hence coincides with the entanglement cost: $\EC(\rho_V)=\EF(\rho_V)=\GEF(V)$ for all two-mode Gaussian states. As a corollary, we obtain in closed form the entanglement cost of simulating the most common noisy bosonic channels. The key inequalities behind these results hold for \emph{every} two-mode state, Gaussian or not: as a consequence, according to both $\EF$ and $\EC$, any two-mode state is at least as entangled as the Gaussian state with the same covariance matrix. This cannot be derived from the general extremality theorem of Ref.~\cite{WolfGiedkeCirac2006}, which applies only to strongly superadditive entanglement measures, a class that excludes the entanglement of formation \cite{Shor2004,Hastings2009}. Experiments routinely use measured second moments to characterize entanglement. Previously, the Gaussian entanglement of formation provided only an upper bound for nonsymmetric Gaussian states. Our results establish it as the \emph{exact} value of both entanglement of formation and entanglement cost for all two-mode Gaussian states and a certified lower bound for any non-Gaussian state with the same covariance matrix.

\textit{Main results.---}
For a two-mode continuous variable system, we collect the canonical quadratures in the vector $R=(x_A,p_A,x_B,p_B)^{\mathsf T}$, satisfying $[R_j,R_k]=i\Omega_{jk}$, where $\Omega=J\oplus J$ is the symplectic form, with $J=\left(\begin{smallmatrix}0&1\\-1&0\end{smallmatrix}\right)$; equivalently, $[x_j,p_k]=i\delta_{jk}$. The annihilation operators are accordingly defined as $a=(x_A+ip_A)/\sqrt2$ and $b=(x_B+ip_B)/\sqrt2$. A state is Gaussian when its phase-space characteristic function, or equivalently its Wigner function, is Gaussian. A Gaussian state is then completely specified by its first moments $d=\langle R\rangle$ and covariance matrix $V$ with entries $V_{jk}=\langle\{R_j-d_j,R_k-d_k\}\rangle/2$, in units where the vacuum has $V=I/2$ \cite{Weedbrook2012,Adesso2014}. Since local displacements do not affect entanglement, we may set the first moments to zero throughout, unless specified otherwise.

For $0<t\leq1$, we define the generalized EPR observable
\begin{align}
 \Dop&=(t a-b^\dagger)^\dagger(t a-b^\dagger)\nonumber\\
 &=\frac12\!\left[(t x_A-x_B)^2+(t p_A+p_B)^2+1-t^2\right].
 \label{eq:Dop}
\end{align}
The parameter $t$ is a relative gain: $t=1$ gives the familiar balanced EPR combination \cite{Duan2000}, while $t<1$ accommodates asymmetric states and channels. For $t<1$, $\Dop$ has a transparent physical meaning:
\begin{equation}
 \Dop=(1-t^2)\left(e_t^\dagger e_t+1\right),\qquad e_t=\frac{b-ta^\dagger}{\sqrt{1-t^2}},
 \label{eq:bogoliubov}
\end{equation}
where $e_t$ is a canonical Bogoliubov mode, $[e_t,e_t^\dagger]=1$. The generalized EPR observable thus counts the excitations of a collective mode of the two parties. Every separable state obeys $\langle\Dop\rangle\geq1$, which makes minimal normalized EPR variances useful inseparability indicators \cite{Duan2000,Marian2018EPR,Marian2021EPR}. The result below turns this qualitative criterion into a quantitative one.

For the two-mode squeezed vacuum (TMSV) state
\begin{equation}
 \TMSV=\sqrt{1-q^2}\sum_{n=0}^{\infty}q^n\lvert n,n\rangle,
 \qquad 0\leq q<1,
 \label{eq:tmsv}
\end{equation}
the entanglement and generalized EPR expectation are
\begin{align}
 E(\Psi_q)=\cE(q)&=g\!\left(\frac{q^2}{1-q^2}\right),
 \label{eq:Eq}\\
 d_t(q)=\langle\Psi_q\rvert\Dop\lvert\Psi_q\rangle
 &=\frac{(1-tq)^2}{1-q^2},
 \label{eq:dtq}
\end{align}
with $g(y)=(y+1)\ln(y+1)-y\ln y$. All logarithms are natural; division by $\ln2$ converts to ebits. Note that $\TMSV$ is the vacuum of $e_t$ precisely when $q=t$.

Our first main result, obtained in consultation with an AI tool, is the following \emph{affine extremality} theorem.

\textit{Theorem 1.---}
For every normalized pure two-mode state $\lvert\psi\rangle$ with finite mean energy and all $0<q<t\leq1$,
\begin{equation}
 E(\psi)+\lambda_{q,t}
 \bigl[\langle\Dop\rangle_\psi-d_t(q)\bigr]
 \geq \cE(q),
 \label{eq:affine}
\end{equation}
where $\lambda_{q,t}=-2q\ln q/[(t-q)(1-tq)]>0$. At $q=t<1$, Eq.~\eqref{eq:affine} holds by continuity: a pure state attaining $d_t(t)$ has entanglement at least $\cE(t)$.

Equation~\eqref{eq:affine} has two complementary meanings. First, at fixed entanglement $\cE(q)$ with $q\leq t$, the TMSV gives the smallest generalized EPR expectation, proving the conjecture isolated in Ref.~\cite{IvanSimon2008}: no state, Gaussian or not, achieves stronger EPR correlations with less entanglement. Second, the inequality is affine in the expectation value of an observable. It can therefore be averaged over an arbitrary ensemble, which is precisely what is needed to control the convex roof. Every finite-energy mixed state, Gaussian or not, obeys
\begin{equation}
 \EF(\rho)\geq
 \sup_{0<q<t\leq1}\!\left\{\cE(q)-\lambda_{q,t}
 \bigl[\Tr(\rho\Dop)-d_t(q)\bigr]\right\}.
 \label{eq:mixedwitness}
\end{equation}
For a centered state, $\Tr(\rho\Dop)$ is obtained from the two variances displayed in Eq.~\eqref{eq:Dop}. Hence Eq.~\eqref{eq:mixedwitness} is a \emph{quantitative witness} of entanglement of formation, accessible with the same phase-resolved records used in standard EPR tests. In the balanced case ($t=1$), writing $\delta=\Tr(\rho\widehat D_1)$, it takes the closed form
\begin{equation}
 \EF(\rho)\geq
 g\!\left(\frac{(1-\delta)^2}{4\delta}\right)
 \quad (0<\delta<1),
 \label{eq:balancedbound}
\end{equation}
with the right-hand side set to zero for $\delta\geq1$.

\textit{Proof strategy.---}
Represent $\lvert\psi\rangle=\sum_{m,n}C_{mn}\lvert m,n\rangle$ by its coefficient matrix $C$, and let $c_0\geq c_1\geq\cdots\geq0$ be the singular values of $C$. Then $c_n^2$ are the Schmidt probabilities, $E(\psi)=H(c^2):=-\sum_n c_n^2\ln c_n^2$, and the EPR expectation reads $\langle\Dop\rangle_\psi=G_t(C)+1-t^2$, with $G_t(C):=\lVert t a^\dagger C-Ca^\dagger\rVert_2^2$. The difficulty is that $G_t$ depends on the Schmidt bases, which for a non-Gaussian state can be arbitrary. The proof proceeds in three steps. First, a new singular-value inequality removes this dependence, bounding the EPR functional by an expression involving only the ordered Schmidt coefficients:
\begin{equation}
 G_t(C)\geq
 \sum_{n=0}^{\infty}(n+1)(t c_n-c_{n+1})_+^2.
 \label{eq:defect}
\end{equation}

Second, whenever $c_{n+1}>t c_n$, we average the offending adjacent blocks after rescaling by the geometric weights $t^{2n}$. This block-averaging produces coefficients $\widetilde c_n$ satisfying $\widetilde c_{n+1}\leq t\widetilde c_n$, while increasing neither the scalar EPR functional nor the Schmidt entropy. It removes precisely the ratio restriction of the earlier rearrangement approach \cite{Akbari2015}.

Third, a logarithmic-Sobolev inequality for phase-covariant Gaussian channels \cite{Beigi2025} shows that the remaining entropy--EPR trade-off is minimized by a geometric sequence:
\begin{equation}
 H(z^2)+\lambda\sum_n(n+1)(tz_n-z_{n+1})^2\geq\inf_{s}\Bigl\{\cE(s)+\lambda\frac{(t-s)^2}{1-s^2}\Bigr\}
 \label{eq:mlsi}
\end{equation}
for every normalized nonnegative sequence $z_n$, where $0\leq s<1$. With $\lambda=\lambda_{q,t}$, the scalar expression on the right has its unique minimum at $s=q$. A geometric sequence $z_n\propto s^n$ is exactly the Schmidt spectrum of a TMSV: the Gaussian optimizer emerges from the argument rather than being assumed. Equations~\eqref{eq:defect} and \eqref{eq:mlsi}, together with the compression step, give Eq.~\eqref{eq:affine}. The Supplemental Material (SM)~\cite{SM}\nocite{PLOB2017,Micuda2012,Hertz2021,Friis2019,Wang2025,Luo2005,DeBievre2019,KoashiWinter2004,Shirokov2023} gives the complete proof, together with the domain and finite-energy limiting arguments required in infinite dimension \cite{Winter2016}.

\textit{Gaussian optimality.---}
We now apply Theorem~1 to Gaussian states. A covariance geometry lemma \cite{IvanSimon2008,Wolf2004}, reproduced in the SM~\cite{SM}, states that every entangled two-mode covariance matrix can be brought by local symplectic transformations (exchanging the parties if necessary) to
\begin{equation}
 V'=V_q+N_{t,u,v},\qquad q\leq t\leq1,
 \label{eq:canonical}
\end{equation}
where $V_q$ is the covariance matrix of $\TMSV$, $u,v\geq0$, and, with $t=\tan\theta$, $c_\theta=\cos\theta$, $s_\theta=\sin\theta$,
\begin{equation}
 N_{t,u,v}=\frac12
 \begin{pmatrix}
 uc_\theta^2&0&uc_\theta s_\theta&0\\
 0&vc_\theta^2&0&-vc_\theta s_\theta\\
 uc_\theta s_\theta&0&us_\theta^2&0\\
 0&-vc_\theta s_\theta&0&vs_\theta^2
 \end{pmatrix}.
 \label{eq:noise}
\end{equation}
Here $q$ labels the smaller-squeezing branch, whose optimality is not assumed.

The added noise has a simple physical meaning. $N_{t,u,v}$ is the covariance of a random local displacement $\xi=(c_\theta X,c_\theta Y,s_\theta X,-s_\theta Y)$, with $X$ and $Y$ independent centered Gaussian variables of variances $u/2$ and $v/2$. Every such displacement moves the two modes in a correlated way that leaves the EPR quadratures $t x_A-x_B$ and $t p_A+p_B$ unchanged. Hence $\rho_{V'}$ is a Gaussian mixture of displaced copies of $\TMSV$, all with the same entanglement $\cE(q)$ and the same EPR expectation $\langle\Dop\rangle=d_t(q)$, component by component and not only on average. In this optimal decomposition, the entanglement is carried by identical squeezed vacua, while the mixedness of $\rho_{V'}$ is entirely due to classical correlated noise. This explains the conditions on the added classical noise found in Ref.~\cite{Marian2008}: $N_{t,u,v}$ is singular, with null directions given precisely by the generalized EPR quadratures, so that the classical state with covariance $I/2+N_{t,u,v}$ sits on the boundary of both classicality and separability~\cite{SM}. This ensemble provides the upper bound $\EF(\rho_{V'})\leq\GEF(V')\leq\cE(q)$.

Conversely, for any pure-state decomposition $\rho_{V'}=\sum_jp_j\lvert\psi_j\rangle\!\langle\psi_j\rvert$, Theorem~1 and linearity of $\Dop$ give
\begin{align}
 \sum_jp_jE(\psi_j)
 &\geq \cE(q)-\lambda_{q,t}
 \left[\Tr(\rho_{V'}\Dop)-d_t(q)\right]
 =\cE(q).
 \label{eq:ensemble}
\end{align}
No ensemble, however non-Gaussian, can do better than the Gaussian one. Local symplectic invariance and the trivial separable case complete the proof of our first main result:
\begin{equation}
 {\EF(\rho_V)=\GEF(V)}.
 \label{eq:main}
\end{equation}
Equation~\eqref{eq:main} holds for every two-mode Gaussian state and resolves the full two-mode instance of \href{https://github.com/google-deepmind/formal-conjectures/issues/3446}{Open Quantum Problem 29} \cite{KruegerWerner2005}.

Our proof turns established results and algorithms for $\GEF$ into exact statements about the {\em unrestricted} $\EF$. These include analytical formulas and bounds for several structured families \cite{AdessoIlluminati2005,Tserkis2017,Wilde2018}, and an efficient one-parameter numerical minimization for a generic two-mode covariance matrix \cite{Tserkis2019}. In covariance matrix form, the entanglement of formation of any two-mode Gaussian state can now be evaluated from the finite-dimensional optimization
\begin{equation}
 \EF(\rho_V)=\min_{\gamma_p\leq V}
 g\!\left(\sqrt{\det\gamma_{p,A}}-\tfrac12\right),
 \label{eq:covopt}
\end{equation}
where $\gamma_p$ ranges over pure two-mode covariance matrices and $\gamma_{p,A}$ is its marginal block for mode $A$; the constraint $\gamma_p\leq V$ expresses that classical displacement noise turns the pure state with covariance $\gamma_p$ into $\rho_V$. The optimization in Eq.~\eqref{eq:covopt} was introduced for $\GEF$ in Ref.~\cite{Wolf2004}; Eq.~\eqref{eq:main} shows that it evaluates the actual $\EF$. In particular, formulas derived in the past under the presumption of Gaussian optimality \cite{Marian2008,Tserkis2017} are now proven exact. A notable example is that of two-mode squeezed thermal states, ubiquitous in experiments, whose covariance matrix has marginal blocks $aI$, $bI$ and correlation block $k\,\mathrm{diag}(1,-1)$. Whenever such a state is entangled, $\EF=g(\sinh^2 r)$, where $r$ is the smaller root of $(a+b)\cosh2r-2k\sinh2r=2(ab-k^2)+\frac12$, which reduces to the symmetric result of Ref.~\cite{Giedke2003} for $a=b$~\cite{SM}.

The equality result also extends to multimode states whose entanglement structure is equivalent to that of two-mode ones. In particular, an $(m+n)$-mode Gaussian state $\rho^{AB}_{\rm bisym}$ is {\em bisymmetric} across $A{:}B$ when its covariance matrix is invariant under arbitrary permutations of the $m$ modes within the $A$ block and of the $n$ modes within the $B$ block. Any bisymmetric $(m+n)$-mode Gaussian state is locally Gaussian-unitarily equivalent to a correlated two-mode core $\rho_{\rm core}$ tensored with $m+n-2$ uncorrelated local thermal modes \cite{Serafini2005}. Adding or removing such local factors changes neither $\EF$ nor $\GEF$, and hence
\begin{equation}
 \EF(\rho^{AB}_{\rm bisym})=\GEF(\rho^{AB}_{\rm bisym})=\EF(\rho_{\rm core}).
 \label{eq:bisym}
\end{equation}

\textit{Additivity and entanglement cost.---}
A natural further question is whether the entanglement of formation is {\em additive}, $\EF(\rho^{\otimes m})=m\EF(\rho)$, and hence equal to the entanglement cost $\EC(\rho)=\lim_{m\to\infty}\frac1m\EF(\rho^{\otimes m})$ \cite{Hayden2001,Yamasaki2025}, for all two-mode Gaussian states. Equation~\eqref{eq:main} alone does not imply this: a collective decomposition of many copies, possibly non-Gaussian, might be cheaper per copy than products of optimal single-copy decompositions. Reference~\cite{Marian2008} claimed additivity for arbitrary two-mode Gaussian states, but its argument only yields the upper bound $\EF(\rho^{\otimes m})\leq m\EF(\rho)$. Additivity was previously known in two special cases: Wolf \emph{et al.} proved it for the \emph{Gaussian} entanglement of formation of symmetric states \cite{Wolf2004}, and Wilde established $\EC=\EF=\GEF$ for states obtained by sending one arm of a TMSV through a quantum-limited attenuator or amplifier \cite{Wilde2018}.

Our second main result settles this question with a many-pair extension of Theorem~1 \footnote{For $m=1$, the proof of Theorem~2~\cite{SM} gives an independent derivation of Theorem~1.}, also obtained in consultation with AI tools. For $m$ pairs of modes $(A_j,B_j)$, with $A=A_1\cdots A_m$ and $B=B_1\cdots B_m$, let $\widehat{D}^{(j)}$ be the observable \eqref{eq:Dop} of the $j$th pair with gain $t_j$ (the roles of $a_j$ and $b_j$ may be exchanged).

\textit{Theorem 2.---}
For every normalized pure state $\lvert\psi\rangle$ of $A{:}B$ with finite mean energy and all $0<q_j<t_j\leq1$,
\begin{equation}
 E(\psi)+\textstyle\sum_j\lambda_{q_j,t_j}\bigl[\langle\widehat{D}^{(j)}\rangle_\psi-d_{t_j}(q_j)\bigr]\geq\sum_j\cE(q_j).
 \label{eq:multiaffine}
\end{equation}

 The proof is variational: at a state minimizing the left-hand side, the second variations along the $2m$ directions $(a_j-q_jb_j^\dagger)\lvert\psi\rangle$ and $(b_j-q_ja_j^\dagger)\lvert\psi\rangle$ must be nonnegative, while by stationarity their sum reduces exactly to a sum of $2\times2$ quadratic forms over pairs of Schmidt coefficients that we prove negative unless all these vectors vanish. The minimizer is thus the joint vacuum $\bigotimes_j\lvert\Psi_{q_j}\rangle$ of these Bogoliubov operators: once more, the Gaussian optimizer emerges rather than being assumed.

\begin{figure*}[t]
 \centering\includegraphics[width=0.8\textwidth]{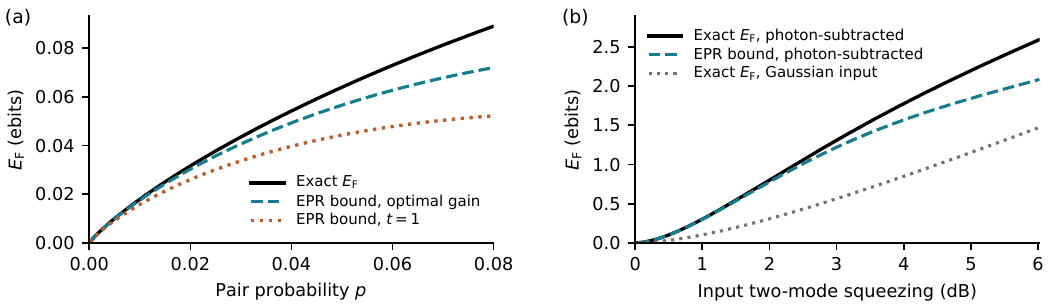}
 \caption{Entanglement of formation of two non-Gaussian states (black solid) and its lower bounds obtained from their covariance matrix alone, which also bound the entanglement cost. (a) Photon-pair state $\sqrt{1-p}\lvert00\rangle+\sqrt p\lvert11\rangle$ sent through pure-loss channels of transmissivities $\eta_A=0.9$ and $\eta_B=0.2$, with exact $\EF$ from Wootters' formula \cite{Wootters1998}. Teal dashed: bound \eqref{eq:covbound}, i.e., Eq.~\eqref{eq:mixedwitness} optimized over the gain $t$; orange dotted: balanced bound \eqref{eq:balancedbound} ($t=1$). (b) Photon-subtracted state $\lvert\Psi_q^-\rangle\propto ab\lvert\Psi_q\rangle$ versus the two-mode squeezing of the original Gaussian input $\lvert\Psi_q\rangle$, $10\log_{10}[(1+q)/(1-q)]$~dB, with exact $\EF=E(\Psi_q^-)$. Teal dashed: bound \eqref{eq:balancedbound}, which coincides with \eqref{eq:covbound} for this symmetric state; grey dotted: entanglement $\cE(q)$ of the Gaussian input.}
 \label{fig:nonGaussian}
\end{figure*}

Averaging Eq.~\eqref{eq:multiaffine} over any decomposition, as in Eq.~\eqref{eq:ensemble}, shows that every state $\omega$ of the $m$ pairs, possibly non-Gaussian and correlated across pairs, obeys
\begin{equation}
 \EF(\omega)\geq\textstyle\sum_j\GEF(V_j),
 \label{eq:superadd}
\end{equation}
where $V_j$ is the covariance matrix of the pair $(A_j,B_j)$. Three consequences follow. First, if $\omega=\sigma_1\otimes\cdots\otimes\sigma_m$ is a product of two-mode Gaussian states, the product of the optimal Gaussian ensembles attains Eq.~\eqref{eq:superadd}. The entanglement of formation is therefore additive on all such products, and in particular
\begin{equation}
 \EC(\rho_V)=\EF(\rho_V)=\GEF(V)
 \label{eq:cost}
\end{equation}
for every two-mode Gaussian state, with the entanglement cost thus given by Eq.~\eqref{eq:covopt}. Second, if the pair marginals $\omega_j$ are Gaussian, Eq.~\eqref{eq:superadd} reads $\EF(\omega)\geq\sum_j\EF(\omega_j)$ for arbitrary $\omega$: on this class, $\EF$ is strongly superadditive, although this property fails in general \cite{Shor2004,Hastings2009}.

Third, Eq.~\eqref{eq:cost} determines the entanglement cost $\EC(\mathcal N)$ of noisy bosonic \emph{channels}: the minimum entanglement per use required for an LOCC simulation with vanishing error, even against adaptive sequential tests \cite{Wilde2018}. For thermal-loss, thermal-amplifier and additive-noise channels, Eq.~\eqref{eq:cost} makes the known lower bounds from finite-energy TMSV outputs meet the upper bounds from teleportation simulation \cite{Wilde2018,TserkisDiasRalph2018}, yielding~\cite{SM}:
\begin{equation}
 \EC(\mathcal N)=
 \begin{cases}
 g\!\left(\dfrac{(1-\delta_{\mathcal N})^2}{4\delta_{\mathcal N}}\right),
       &0<\delta_{\mathcal N}<1,\\[2pt]
 0,    &\delta_{\mathcal N}\geq1 .
 \end{cases}
 \label{eq:channelcost}
\end{equation}
Here $\delta_{\mathcal N}=(\sqrt{N_B+1}+\sqrt{N_B})^2\,|1-\sqrt\tau|/(1+\sqrt\tau)$ for transmissivity or gain $\tau$ and $N_B$ thermal photons in the environment, and $\delta_{\mathcal N}=\xi$ for additive noise of variance $\xi$. The zero-cost branch corresponds to entanglement-breaking channels; the identity-channel limit $\delta_{\mathcal N}=0$ has divergent cost. Costs are in nats per use, with division by $\ln2$ giving the Bell-pair rate.

\textit{Non-Gaussian states and experimental applications.---}
Equation~\eqref{eq:mixedwitness} does not assume Gaussianity, and hence bounds the entanglement of formation of {\em any} two-mode state. Its strongest form is obtained in the local frame where the covariance matrix $V$ of $\rho$ takes the canonical form \eqref{eq:canonical}: since $\Tr(\rho\Dop)$ depends only on $V$, the arguments leading to Eqs.~\eqref{eq:main} and \eqref{eq:superadd} go through verbatim and give~\cite{SM}
\begin{equation}
 \EF(\rho)\geq\EC(\rho)\geq\GEF(V).
 \label{eq:covbound}
\end{equation}
Hence, according to both the entanglement of formation and the entanglement cost, a two-mode state can never be less entangled than the Gaussian state with the same first and second moments. This cannot be obtained from the extremality theorem of Wolf, Giedke and Cirac \cite{WolfGiedkeCirac2006}, which requires the entanglement measure to be strongly superadditive on all states, a property violated in general by the entanglement of formation \cite{Shor2004,Hastings2009}. Such certification is not automatic: logarithmic negativity can exceed its true value when evaluated on the Gaussian state with the same covariance matrix \cite{WolfGiedkeCirac2006}. Equation~\eqref{eq:covopt}, however, gives a tight lower bound on both $\EF$ and $\EC$ without requiring Gaussianity to be verified.

Figure~\ref{fig:nonGaussian} illustrates Eq.~\eqref{eq:covbound} for two non-Gaussian states of current experimental interest. The first models a weak photon-pair source as $\sqrt{1-p}\lvert00\rangle+\sqrt p\lvert11\rangle$, followed by losses with unequal transmissivities. Each mode remains in the zero- and one-photon subspace, allowing the exact $\EF$ to be evaluated using Wootters' formula for two-qubit states \cite{Wootters1998}. Tuning the gain $t$ compensates for the loss asymmetry: at $p=0.01$, the optimized bound \eqref{eq:covbound} captures $98\%$ of the exact value, compared with $88\%$ for the balanced measurement $t=1$. 

The second is a genuinely continuous variable state, obtained from a TMSV by removing one photon from each mode. Such photon subtraction, implemented by heralded detection of a small fraction of each beam tapped off on a weakly reflecting beam splitter, is a standard route to the distillation of continuous variable entanglement \cite{Ourjoumtsev2007,Kurochkin2014}. The resulting state is pure but non-Gaussian, so that its exact $\EF$ is its entanglement entropy. Photon subtraction enhances the EPR correlations at every squeezing level, and Eq.~\eqref{eq:balancedbound} converts this enhancement into a certified increase of the entanglement of formation over the original Gaussian input, capturing at least $80\%$ of the exact value over the range shown. The certified increase persists under equal losses~\cite{SM}.

Experimentally, the tight bound \eqref{eq:covbound} follows from the measured covariance matrix, while two suitably chosen EPR variances already provide the quantitative witness \eqref{eq:mixedwitness}. In optics, two-mode covariance matrices are routinely reconstructed by balanced homodyne detection \cite{Laurat2005,Jia2025}. For propagating microwave fields in superconducting circuits, the quadratures are recorded by linear amplification followed by heterodyne digitization, and the covariance matrix is obtained after subtracting the calibrated amplifier noise \cite{Menzel2012}; the same approach characterizes entangled microwave--optical fields from electro-optic transducers \cite{Sahu2023}. A single data set serves for any gain $t$, with no photon-number-resolving detection or non-Gaussian tomography; calibration and finite statistics are discussed in the SM~\cite{SM}. Finally, when the covariance matrix of a non-Gaussian state hides its entanglement, a probabilistic local operation (filtering) before the quadrature measurement may reveal it: since neither the entanglement of formation nor the entanglement cost can increase on average under local operations, the bound on the filtered state, times the success probability, certifies both quantities for the original state~\cite{SM}.

\textit{Conclusion and Outlook.---}
We have proven Gaussian optimality and additivity of the entanglement of formation for all two-mode Gaussian states, resolving the full two-mode instance of \href{https://github.com/google-deepmind/formal-conjectures/issues/3446}{Open Quantum Problem 29} \cite{KruegerWerner2005} and determining the exact entanglement cost of this pivotal family of continuous variable states. Collective non-Gaussian preparations cannot reduce the entanglement required per copy. The same result determines the cost of simulating thermal-loss, thermal-amplifier and additive-noise channels, which play a central role in optical communication \cite{Weedbrook2012}.

The link to EPR correlations makes these results useful beyond Gaussian sources. Measured covariance matrices give certified lower bounds on both entanglement of formation and entanglement cost for arbitrary two-mode states, with equality for Gaussian states. This connects the resources required for quantum communication to quadrature measurements already available in optical and microwave experiments, without requiring full state tomography or verification of Gaussianity.

Our proofs identify the Gaussian optimizers through complementary methods: reduction to Schmidt spectra (Theorem~1) and stability of variational minimizers (Theorem~2). Beyond entanglement, our results also connect single-mode entropy to the quantum contribution to quadrature fluctuations, quantified by the Wigner--Yanase skew information \cite{WignerYanase1963,SM}. They further establish the optimality of Gaussian measurements for quantum discord \cite{OllivierZurek2001,HendersonVedral2001,AdessoDatta2010,GiordaParis2010,Pirandola2014}, including its regularization, in two-mode Gaussian states with one vacuum normal mode~\cite{SM}. Extending these results to general multimode Gaussian states, beyond those reducible to two-mode cores, remains an open problem. Finally, this work illustrates the emerging role of large language models as research assistants in mathematical physics \cite{AImagazine}.

\begin{acknowledgments}
\textit{Acknowledgments.---}
I acknowledge substantive use of AI tools to assist with research and development of results presented here. Through iterative exchanges in which I proposed, tested and rejected candidate routes, ChatGPT (GPT 5.6 Sol by OpenAI) contributed to the affine supporting-line formulation of Theorem~1, to the geometric isotonic compression, and to identifying the logarithmic-Sobolev inequality of Ref.~\cite{Beigi2025} as the closing step; it also assisted with algebraic derivations, numerical checks, literature organization, and drafting. The proof of Theorem~2 was developed with assistance from ChatGPT (GPT 6 Astra by OpenAI) and independently verified with Claude (Opus 5.5 by Anthropic), which also assisted with other results in the revised version. I directed the research, checked every step of the proofs and every scientific claim, and take full responsibility for this work. I thank Mark Wilde for pointing out Ref.~\cite{Wilde2018}. This work was supported by the Engineering and Physical Sciences Research Council (EPSRC) Grant No.~UKRI6536.
\end{acknowledgments}

\bibliography{references}


\onecolumngrid
\clearpage




\setcounter{section}{0}
\setcounter{subsection}{0}
\setcounter{equation}{0}
\setcounter{figure}{0}
\setcounter{table}{0}

\setcounter{secnumdepth}{1}
\renewcommand{\thesection}{S\arabic{section}}
\renewcommand{\thesubsection}{S\arabic{section}.\arabic{subsection}}
\renewcommand{\theequation}{S\arabic{equation}}
\renewcommand{\thefigure}{S\arabic{figure}}
\renewcommand{\thetable}{S\arabic{table}}
\renewcommand{\theHequation}{S.\arabic{equation}}
\renewcommand{\theHsection}{S.\arabic{section}}
\renewcommand{\theHsubsection}{S.\arabic{section}.\arabic{subsection}}
\theoremstyle{plain}
\newtheorem{theorem}{Theorem}
\renewcommand{\thetheorem}{S\arabic{theorem}}
\newtheorem{lemma}[theorem]{Lemma}
\newtheorem{corollary}[theorem]{Corollary}

\onecolumngrid
\begin{center}
{\large\bfseries Supplemental Material for\\[2pt]
``Optimality of Gaussian Entanglement of Formation''}
\end{center}
\vspace{1em}
\twocolumngrid

\section{Preliminaries}

We use throughout the notation and conventions of the Letter.
A normalized pure two-mode state is represented by a Hilbert--Schmidt
coefficient operator $C$ according to
\begin{equation}
 \lvert\psi_C\rangle
 =\sum_{m,n\geq0}C_{mn}\lvert m,n\rangle,
 \qquad
 \lVert C\rVert_2=1.
 \label{S:Cstate}
\end{equation}
If
$c_0\geq c_1\geq\cdots\geq0$
are the singular values of $C$, then $p_n=c_n^2$ are the Schmidt
probabilities and
\begin{equation}
 E(\psi_C)=H(c^2):=-\sum_n c_n^2\ln c_n^2.
 \label{S:entropy}
\end{equation}
We use the vectorization identities
\begin{equation}
 (A\otimes I)\lvert C\rangle\!\rangle=\lvert AC\rangle\!\rangle,
 \qquad
 (I\otimes B)\lvert C\rangle\!\rangle
 =\lvert CB^{\mathsf T}\rangle\!\rangle.
 \label{S:vec}
\end{equation}
Since the Fock-basis matrix of the annihilation operator satisfies
$a^{\mathsf T}=a^\dagger$, the generalized EPR expectation introduced
in the Letter can be written as
\begin{align}
 D_t(\psi_C)
 &:=\langle\psi_C\rvert\widehat D_t\lvert\psi_C\rangle,
 \nonumber\\
 G_t(C)
 &:=\lVert t a^\dagger C-Ca^\dagger\rVert_2^2
 =D_t(\psi_C)-1+t^2 .
 \label{S:Gdef}
\end{align}

We record here the domain facts needed below.
For a finite-energy pure state,
\begin{equation}
 \sum_k c_k^2\langle u_k|N|u_k\rangle<\infty,
 \qquad
 \sum_k c_k^2\langle v_k|N|v_k\rangle<\infty,
 \label{S:localenergy}
\end{equation}
where $\{|u_k\rangle\}$ and $\{|v_k\rangle\}$ are left and right
Schmidt vectors and $N=a^\dagger a$.

Since all terms in Eq.~\eqref{S:localenergy} are nonnegative, every
Schmidt vector with $c_k>0$ has finite number expectation. Hence
$|u_k\rangle$ and $|v_k\rangle$ belong to
$\operatorname{Dom}(a)\cap\operatorname{Dom}(a^\dagger)$ whenever
$c_k>0$. Moreover, if either of the dual potentials $\alpha_i,\beta_i$
introduced below is nonzero, then necessarily $c_i>0$. Therefore, for
every index that contributes to the dual potential sums, completeness
of the Schmidt bases gives
\begin{align}
 \sum_j|X_{ij}|^2&=\|a^\dagger u_i\|^2,
 &
 \sum_j|X_{ji}|^2&=\|a u_i\|^2,
 \nonumber\\
 \sum_j|Y_{ij}|^2&=\|a^\dagger v_i\|^2,
 &
 \sum_j|Y_{ji}|^2&=\|a v_i\|^2 .
 \label{S:domainrows}
\end{align}
The canonical commutation relation therefore gives the index identities
used below for every index that contributes to the dual potential sums.

The passive rearrangement principle furthermore gives
\begin{equation}
 \sum_n n c_n^2
 \leq \Tr(\rho_A a^\dagger a)<\infty.
 \label{S:passiveenergy}
\end{equation}
This fact will be used when passing from finite Schmidt rank to the
general finite-energy case.

\section{Proof of Theorem~1}
\label{S:thm1proof}

The proof consists of the three steps outlined in the Letter:
a singular-value defect inequality, a geometric isotonic compression,
and a thermal log-Sobolev bound. We give the complete arguments in
turn.

\subsection{Singular-value defect inequality}

\begin{theorem}\label{thm:defect}
Let $0<t\leq1$, and let $C$ be a normalized Hilbert--Schmidt operator
with finite local mean energies. If $c_n$ are its decreasing singular
values, then
\begin{equation}
 G_t(C)\geq\sum_{n=0}^\infty(n+1)d_n^2,
 \qquad
 d_n=(t c_n-c_{n+1})_+.
 \label{S:defect}
\end{equation}
\end{theorem}

\begin{proof}
We first establish a positive-drop lemma. For $i<j$, define
\begin{equation}
 D_{ij}=\sum_{k=i}^{j-1}(t c_k-c_{k+1})_+ .
 \label{S:Dij}
\end{equation}
Then
\begin{equation}
 D_{ij}\leq(t c_i-c_j)_+ .
 \label{S:drop}
\end{equation}
To prove this, decompose the active indices into maximal consecutive
intervals $[\ell,r]$. On each interval,
\begin{align}
 \sum_{k=\ell}^r(t c_k-c_{k+1})
 &=t c_\ell-c_{r+1}
 -(1-t)\sum_{k=\ell+1}^r c_k
 \nonumber\\
 &\leq t c_\ell-c_{r+1}.
 \label{S:blockdrop}
\end{align}
For two successive active intervals, monotonicity gives
$t c_{\ell_2}\leq c_{r_1+1}$, so their bounds concatenate.
Iteration proves Eq.~\eqref{S:drop}. The same statement, with the
endpoints relabeled, will be used when $i>j$.

We now prove the theorem, first for finite Schmidt rank. Both kernels are
then infinite dimensional, so the singular vectors can be completed to
paired orthonormal bases of the full local spaces. Take a singular-value
decomposition
$C=U\,\mathrm{diag}(c)\,V^\dagger$ and set
\begin{equation}
 X=U^\dagger aU,
 \qquad
 Y=V^\dagger aV.
 \label{S:XY}
\end{equation}
After relabeling matrix indices,
\begin{equation}
 G_t(C)
 =\sum_{i,j\geq0}
 \lvert t c_iX_{ij}-c_jY_{ij}\rvert^2.
 \label{S:Gmatrix}
\end{equation}

Using Eq.~\eqref{S:domainrows} and
$aa^\dagger-a^\dagger a=I$, we obtain, for every relevant index,
\begin{equation}
 \sum_j\lvert X_{ij}\rvert^2-\sum_j\lvert X_{ji}\rvert^2=1,
 \qquad
 \sum_j\lvert Y_{ij}\rvert^2-\sum_j\lvert Y_{ji}\rvert^2=1.
 \label{S:index}
\end{equation}

Fix a cutoff $N$, put $\delta_k=d_k$ for $k\leq N$ and
$\delta_k=0$ for $k>N$, and define
\begin{equation}
 \alpha_i=\sum_{k=i}^{\infty} t c_k\delta_k,
 \qquad
 \beta_i=-\sum_{k=i}^{\infty}c_{k+1}\delta_k .
 \label{S:potentials}
\end{equation}
Thus $\alpha_i=\beta_i=0$ for $i>N$. The truncation ensures that the
dual potentials have finite support, so that the weighted row and column
sums below converge absolutely; at the end $N$ is sent to infinity. We claim that, for arbitrary
$x,y\in\mathbb C$,
\begin{equation}
 \lvert t c_ix-c_jy\rvert^2
 \geq(\alpha_i-\alpha_j)\lvert x\rvert^2
 +(\beta_i-\beta_j)\lvert y\rvert^2.
 \label{S:pointwise}
\end{equation}
This is equivalent to positivity of
\begin{equation}
 M_{ij}=
 \begin{pmatrix}
  t^2c_i^2-(\alpha_i-\alpha_j)&-tc_ic_j\\
  -tc_ic_j&c_j^2-(\beta_i-\beta_j)
 \end{pmatrix}.
 \label{S:Mij}
\end{equation}

Consider first $i<j$ and put
\begin{equation}
 D=\sum_{k=i}^{j-1}\delta_k,
 \qquad
 P=\sum_{k=i}^{j-1}tc_k\delta_k,
 \qquad
 Q=\sum_{k=i}^{j-1}c_{k+1}\delta_k.
 \label{S:DPQ}
\end{equation}
If $D=0$, then $P=Q=0$ and $M_{ij}$ is the rank-one positive
matrix generated by $(tc_i,-c_j)$. If $D>0$,
Eq.~\eqref{S:drop} implies $D\leq tc_i-c_j$, while monotonicity gives
\begin{equation}
 P\leq tc_iD,
 \qquad
 Q\geq c_jD.
 \label{S:PQ1}
\end{equation}
Writing $A=t^2c_i^2-P$, we obtain
\begin{equation}
 A\geq tc_i(tc_i-D)\geq tc_ic_j.
 \label{S:Abound}
\end{equation}
The determinant satisfies
\begin{equation}
 \det M_{ij}=AQ-Pc_j^2\geq0
 \label{S:det1}
\end{equation}
by Eq.~\eqref{S:PQ1}. Hence $M_{ij}\geq0$.

For $i>j$, define $D,P,Q$ over $k=j,\ldots,i-1$.
Again the case $D=0$ is immediate. Otherwise,
\begin{equation}
 D\leq tc_j-c_i,
 \qquad
 P\geq tc_iD,
 \qquad
 Q\leq c_jD.
 \label{S:PQ2}
\end{equation}
Now $M_{ij}$ has diagonal entries
$A=t^2c_i^2+P$ and $B=c_j^2-Q$.
Equation~\eqref{S:PQ2} yields
\begin{equation}
 B\geq c_j(c_j-D)\geq c_jc_i.
 \label{S:Bbound}
\end{equation}
Consequently,
\begin{align}
 PB&\geq t c_i^2c_jD
 \geq t^2c_i^2c_jD
 \geq t^2c_i^2Q,
 \nonumber\\
 \det M_{ij}&=PB-t^2c_i^2Q\geq0 .
 \label{S:det2}
\end{align}
The case $i=j$ is immediate, proving Eq.~\eqref{S:pointwise}.

Apply Eq.~\eqref{S:pointwise} to Eq.~\eqref{S:Gmatrix}.
Since the potentials have finite support, Eq.~\eqref{S:index} may be
used without an exchange of divergent sums:
\begin{align}
 G_t(C)
 &\geq\sum_i(\alpha_i+\beta_i)
 \nonumber\\
 &=\sum_{k=0}^N(k+1)(tc_k-c_{k+1})d_k
 \nonumber\\
 &=\sum_{k=0}^N(k+1)d_k^2 .
 \label{S:defectN}
\end{align}
Monotone convergence as $N\to\infty$ proves the theorem for finite
Schmidt rank. For general $C$, let $C^{(L)}$ be its normalized Schmidt
truncation to the $L+1$ largest singular values. Finite local energy gives
$a^\dagger C^{(L)}\to a^\dagger C$ and $C^{(L)}a^\dagger\to Ca^\dagger$ in
Hilbert--Schmidt norm, hence $G_t(C^{(L)})\to G_t(C)$. Applying the
finite-rank inequality to $C^{(L)}$, keeping the terms $n\leq N$ for fixed
$N$, and letting first $L\to\infty$ and then $N\to\infty$ concludes the
proof.
\end{proof}

\subsection{Geometric isotonic compression}

\begin{theorem}\label{thm:compression}
Let
$c_0\geq c_1\geq\cdots\geq c_M\geq0$,
$\sum_{n=0}^M c_n^2=1$, set $c_{M+1}=0$, and let $0<t<1$.
There exists a normalized sequence
$\widetilde c_0,\ldots,\widetilde c_M\geq0$ such that
\begin{align}
 \widetilde c_{n+1}
 &\leq t\widetilde c_n,
 \label{S:ratio}\\
 H(\widetilde c^2)
 &\leq H(c^2),
 \label{S:entropycomp}\\
 \sum_n(n+1)(t\widetilde c_n-\widetilde c_{n+1})^2
 &\leq
 \sum_n(n+1)(tc_n-c_{n+1})_+^2.
 \label{S:energycomp}
\end{align}
For $t=1$, take $\widetilde c=c$.
All sums in this subsection are initially over $0\leq n\leq M$,
with the boundary value at $M+1$ fixed to zero.
\end{theorem}

\begin{proof}
Set
\begin{equation}
 p_n=c_n^2,
 \qquad
 w_n=t^{2n},
 \qquad
 r_n=p_n/w_n .
 \label{S:rdef}
\end{equation}
Perform weighted antitonic regression of $r_n$ with weights $w_n$.
Equivalently, apply the pool-adjacent-violators algorithm:
neighboring constant blocks with levels $r_L<r_R$ are replaced by
their weighted mean
\begin{equation}
 r_{\mathrm{new}}
 =\frac{W_Lr_L+W_Rr_R}{W_L+W_R},
 \label{S:pool}
\end{equation}
where $W_L,W_R$ are the sums of $w_n$ over the blocks.
Let $\widetilde r_n$ be the final nonincreasing fit and define
\begin{equation}
 \widetilde p_n=w_n\widetilde r_n,
 \qquad
 \widetilde c_n=\sqrt{\widetilde p_n}.
 \label{S:tildec}
\end{equation}
Normalization is preserved blockwise. Since
$\widetilde r_{n+1}\leq\widetilde r_n$,
\begin{equation}
 \widetilde p_{n+1}
 =t^{2n+2}\widetilde r_{n+1}
 \leq t^2\widetilde p_n,
 \label{S:ratioP}
\end{equation}
which proves Eq.~\eqref{S:ratio}, including when one or both fitted
levels vanish.

To prove Eq.~\eqref{S:energycomp}, write $s_n=\sqrt{r_n}$. Then
\begin{equation}
 \sum_n(n+1)(tc_n-c_{n+1})_+^2
 =
 \sum_n(n+1)t^{2n+2}(s_n-s_{n+1})_+^2 .
 \label{S:senergy}
\end{equation}
Pooling adjacent block levels $x<y$ replaces their square roots by
a common value between $\sqrt{x}$ and $\sqrt{y}$.
The internal boundary contributes zero both before and after pooling.
The left external positive drop cannot increase because the left block
level is raised, while the right external positive drop cannot
increase because the right block level is lowered.
Thus every pooling operation decreases Eq.~\eqref{S:senergy}.
At termination, all drops are nonnegative without the positive-part
operation, proving Eq.~\eqref{S:energycomp}.

For entropy, introduce cumulative weights and masses,
\begin{equation}
 W_K=\sum_{n=0}^K w_n,
 \qquad
 P_K=\sum_{n=0}^K p_n.
 \label{S:WP}
\end{equation}
Weighted antitonic regression is equivalently obtained by taking the
least concave majorant of the polygonal line through $(0,0)$ and
$(W_K,P_K)$; the fitted levels $\widetilde r_n$ are its successive
slopes. The majorant lies above the original line and has the same
final endpoint. Therefore, for every $K$,
\begin{equation}
 \sum_{n=0}^K\widetilde p_n
 \geq
 \sum_{n=0}^Kp_n,
 \label{S:prefix}
\end{equation}
with equality at the end of every final block produced by the
pool-adjacent-violators algorithm.
Both $p$ and $\widetilde p$ are decreasing sequences; hence
$\widetilde p$ majorizes $p$.
Shannon entropy is Schur concave, proving
Eq.~\eqref{S:entropycomp}.
Majorization also gives
$\sum_n n\widetilde p_n\leq\sum_nnp_n$.
\end{proof}

\subsection{Thermal log-Sobolev closure}

We now derive the scalar Gaussian-optimizer inequality used in the
Letter. Let $z_n\geq0$, $\sum_n z_n^2=1$, and define the diagonal state
\begin{equation}
 \rho_z=\sum_n z_n^2\lvert n\rangle\!\langle n\rvert .
 \label{S:rhoz}
\end{equation}
Set
\begin{equation}
 \cF_t(z)
 =\sum_{n=0}^\infty(n+1)(tz_n-z_{n+1})^2 .
 \label{S:Fdef}
\end{equation}
For
$\bar n=\Tr(\rho_z a^\dagger a)$ and
\begin{equation}
 T_z
 =\Tr(\sqrt{\rho_z}\,a\sqrt{\rho_z}\,a^\dagger)
 =\sum_n(n+1)z_nz_{n+1},
 \label{S:Tz}
\end{equation}
we have
\begin{equation}
 \cF_t(z)
 =t^2(\bar n+1)+\bar n-2tT_z .
 \label{S:Fmoments}
\end{equation}

The meta logarithmic-Sobolev theorem of Beigi and Rahimi-Keshari
\cite{Beigi2025} applies to
\begin{align}
 \Upsilon(\rho)
 ={}&
 2\nu_0\left[
 \Tr(\rho aa^\dagger)
 -\Tr(\sqrt\rho\,a\sqrt\rho\,a^\dagger)
 \right]
 \nonumber\\
 &+
 2\nu_1\left[
 \Tr(\rho a^\dagger a)
 -\Tr(\sqrt\rho\,a^\dagger\sqrt\rho\,a)
 \right]
 \nonumber\\
 &+\omega\Tr(\rho a^\dagger a)+S(\rho)
 \label{S:Upsilon}
\end{align}
for arbitrary $\nu_0,\nu_1,\omega\geq0$, and states that its infimum
is attained among thermal states. Choose
\begin{equation}
 \nu_0=\lambda t,
 \qquad
 \nu_1=0,
 \qquad
 \omega=\lambda(1-t)^2 .
 \label{S:parameters}
\end{equation}
Using Eq.~\eqref{S:Fmoments},
\begin{equation}
\Upsilon(\rho_z)
=H(z^2)+\lambda\cF_t(z)+\lambda(2t-t^2).
 \label{S:UpsilonF}
\end{equation}
The additive constant is state independent.
Thermal states can be parameterized as
\begin{equation}
 \tau_s=(1-s^2)\sum_{n=0}^\infty
 s^{2n}\lvert n\rangle\!\langle n\rvert,
 \qquad
 0\leq s<1.
 \label{S:taus}
\end{equation}
Define the corresponding square-root eigenvalue sequence by
\begin{equation}
 z_n^{(s)}=\sqrt{1-s^2}\,s^n .
 \label{S:thermalz}
\end{equation}
Then
\begin{equation}
 S(\tau_s)=\cE(s),
 \qquad
 \cF_t(z^{(s)})
 =\frac{(t-s)^2}{1-s^2}.
 \label{S:thermalF}
\end{equation}
Therefore
\begin{equation}
 H(z^2)+\lambda\cF_t(z)
 \geq
 \inf_{0\leq s<1}
 \left[
 \cE(s)+\lambda\frac{(t-s)^2}{1-s^2}
 \right].
 \label{S:thermalopt}
\end{equation}

For $0<q<t$, choose
\begin{equation}
 \lambda_{q,t}
 =\frac{-2q\ln q}{(t-q)(1-tq)} .
 \label{S:lambdadef}
\end{equation}
Indeed,
\begin{align}
 \cE'(s)
 &=\frac{-4s\ln s}{(1-s^2)^2},
 \label{S:Eprime}\\
 \frac{d}{ds}\frac{(t-s)^2}{1-s^2}
 &=
 -\frac{2(t-s)(1-ts)}{(1-s^2)^2},
 \label{S:Fprime}
\end{align}
so $s=q$ is stationary.
To see that this stationary point is the unique global minimizer,
define
\begin{equation}
 \Lambda_t(s)
 =\frac{-2s\ln s}{(t-s)(1-ts)},
 \qquad 0<s<t,
 \label{S:Lambdadef}
\end{equation}
so that $\lambda_{q,t}=\Lambda_t(q)$.
Then
\begin{equation}
 \frac{d}{ds}\ln\Lambda_t(s)
 =
 \frac1s+\frac{1}{s\ln s}
 +\frac1{t-s}+\frac{t}{1-ts}>0 .
 \label{S:LambdaMon}
\end{equation}
Indeed, the first three terms on the right-hand side of
Eq.~\eqref{S:LambdaMon} are bounded below by
\begin{equation}
 \frac1s+\frac{1}{s\ln s}+\frac1{1-s}
 =
 \frac{-\ln s-(1-s)}{-s(1-s)\ln s}>0,
 \label{S:LambdaBound}
\end{equation}
and the last term is positive.
Thus $\Lambda_t$ is strictly increasing.
For $\lambda=\lambda_{q,t}=\Lambda_t(q)$, the scalar objective decreases for
$s<q$, increases for $q<s<t$, and is manifestly increasing for
$s\geq t$. Consequently,
\begin{equation}
 H(z^2)+\lambda_{q,t}\cF_t(z)
 \geq
 \cE(q)
 +\lambda_{q,t}\frac{(t-q)^2}{1-q^2}.
 \label{S:support}
\end{equation}
Equation~\eqref{S:support} is a global supporting-line inequality for
the thermal entropy--energy curve.

\subsection{Completion of the proof and infinite-dimensional limit}

We first justify the passage from the finite-rank compression of
Theorem~\ref{thm:compression} to a general finite-energy Schmidt spectrum.
Let $c^{(M)}$ denote the normalized truncation of $c$ to
$n=0,\ldots,M$, and let $\widetilde c^{(M)}$ be the sequence obtained
from Theorem~\ref{thm:compression}, with
$\widetilde p_n^{(M)}=(\widetilde c_n^{(M)})^2$.
By Eq.~\eqref{S:passiveenergy}, the normalized truncations converge in
trace norm and have uniformly bounded mean photon number.
Energy-constrained continuity of the von Neumann entropy therefore gives \cite{Winter2016}
\begin{equation}
 H\!\left((c^{(M)})^2\right)\longrightarrow H(c^2) .
 \label{S:entropytrunc}
\end{equation}

Moreover, Theorem~\ref{thm:compression} gives
\begin{equation}
 \sum_n n\,\widetilde p_n^{(M)}
 \leq
 \sum_n n\,(c_n^{(M)})^2 ,
 \label{S:tight}
\end{equation}
so $\{\widetilde p^{(M)}\}_M$ has uniformly bounded first moment and is
therefore tight. Hence, along a subsequence,
$\widetilde p^{(M)}\to\widetilde p$ in $\ell^1$ for some normalized
probability distribution $\widetilde p$. Setting
$\widetilde c_n=\sqrt{\widetilde p_n}$, the inequality
$\widetilde c_{n+1}^{(M)}\leq t\widetilde c_n^{(M)}$ passes to the limit.
Lower semicontinuity of the entropy gives
\begin{equation}
 H(\widetilde c^2)
 \leq
 \liminf_{M\to\infty}H\!\left((\widetilde c^{(M)})^2\right)
 \leq H(c^2).
 \label{S:entropyinf}
\end{equation}
Likewise, lower semicontinuity of the nonnegative Dirichlet sum yields
\begin{align}
 \sum_n(n+1)(t\widetilde c_n-\widetilde c_{n+1})^2
 &\leq
 \liminf_{M\to\infty}
 \sum_n(n+1)
 (t\widetilde c_n^{(M)}-\widetilde c_{n+1}^{(M)})^2
 \nonumber\\
 &\leq
 \sum_n(n+1)(t c_n-c_{n+1})_+^2 .
 \label{S:energyinf}
\end{align}
In the last step we used Eq.~\eqref{S:energycomp} and convergence of the
truncated defect sums, which follows from Eq.~\eqref{S:passiveenergy}.
Thus Theorem~\ref{thm:compression} extends to arbitrary finite-energy Schmidt spectra.

We may now combine Theorems~\ref{thm:defect} and \ref{thm:compression} with
Eq.~\eqref{S:support}. For an arbitrary finite-energy coefficient
operator $C$ with singular values $c$,
\begin{align}
 E(\psi_C)+\lambda_{q,t}G_t(C)
 &\geq
 H(\widetilde c^2)
 +\lambda_{q,t}\cF_t(\widetilde c)
 \nonumber\\
 &\geq
 \cE(q)
 +\lambda_{q,t}\frac{(t-q)^2}{1-q^2}.
 \label{S:affineG}
\end{align}
Using Eq.~\eqref{S:Gdef} and the identity
\begin{equation}
 d_t(q)-1+t^2
 =\frac{(t-q)^2}{1-q^2},
 \label{S:Gtq}
\end{equation}
which follows directly from the formula $d_t(q)=(1-tq)^2/(1-q^2)$ in the Letter,
Eq.~\eqref{S:affineG} becomes
\begin{equation}
 E(\psi_C)
 +\lambda_{q,t}
 \bigl[D_t(\psi_C)-d_t(q)\bigr]
 \geq\cE(q).
 \label{S:affineD}
\end{equation}
This proves Theorem~1.

The endpoint $q=t<1$ is obtained by taking $q'\uparrow t$ in
Eq.~\eqref{S:affineD}. In particular, if
$D_t(\psi)=d_t(t)=1-t^2$, then
\begin{equation}
 E(\psi)
 \geq
 \cE(q')
 +\lambda_{q',t}
 \bigl[d_t(q')-d_t(t)\bigr]
 \longrightarrow
 \cE(t).
 \label{S:endpoint}
\end{equation}
The case $q=0$ is the separable boundary.

The affine bound also holds for any finite-energy pure state $\lvert\psi\rangle$ of a larger bipartite bosonic system, with $a$ and $b$ any single modes on the two sides. Indeed, tracing out all other modes is a local operation, so $E(\psi)\geq E_{\mathrm F}(\rho_{ab})$. Applying the mixed-state bound obtained by averaging Theorem~1 to $\rho_{ab}$ gives
\begin{equation}
 E(\psi)\geq E_{\mathrm F}(\rho_{ab})
 \geq\mathcal E(q)-\lambda_{q,t}
 \bigl[\langle\widehat D_t\rangle_\psi-d_t(q)\bigr].
\end{equation}
Thus the extension follows from LOCC monotonicity without repeating the Schmidt-basis argument in the larger local spaces.

The affine theorem also gives directly the fixed-entanglement
extremality statement used in the Letter:
\begin{equation}
 E(\psi)\leq\cE(q)
 \quad\Longrightarrow\quad
 D_t(\psi)\geq d_t(q),
 \qquad q\leq t .
 \label{S:fixedE}
\end{equation}
For general convex decompositions represented by probability
measures over pure states, Eq.~\eqref{S:affineD} may be integrated
and Tonelli's theorem applied to the nonnegative energy terms.
This gives the ensemble-averaged affine bound used in the Letter
without requiring a countable decomposition.

\section{Canonical covariance decomposition}

We now recall the purely Gaussian covariance result used in the
proof of $E_{\mathrm F}=E_{\mathrm F}^{\mathrm G}$ in the Letter. Canonical quadratures  are
ordered as $(x_A,p_A,x_B,p_B)$ and, with the conventions of the
Letter, the vacuum covariance is $I_4/2$.

\begin{lemma}[Two-mode canonical form]\label{lem:canonical}
For every entangled two-mode covariance matrix $V$, after exchanging
the parties if necessary, there exist local
symplectic matrices $S_A,S_B$, parameters
$0<q\leq t\leq1$, and $u,v\geq0$ such that
\begin{equation}
 (S_A\oplus S_B)V(S_A\oplus S_B)^{\mathsf T}
 =V_q+N_{t,u,v},
 \label{S:canonical}
\end{equation}
where
\begin{equation}
 V_q=
 \frac{1}{2(1-q^2)}
 \begin{pmatrix}
  1+q^2&0&2q&0\\
  0&1+q^2&0&-2q\\
  2q&0&1+q^2&0\\
  0&-2q&0&1+q^2
 \end{pmatrix},
 \label{S:Vq}
\end{equation}
and, with $t=\tan\theta$, $c_\theta=\cos\theta$, $s_\theta=\sin\theta$, $0<\theta\leq\pi/4$,
\begin{equation}
 N_{t,u,v}
 =
 \frac12
 \begin{pmatrix}
  uc_\theta^2&0&uc_\theta s_\theta&0\\
  0&vc_\theta^2&0&-vc_\theta s_\theta\\
  uc_\theta s_\theta&0&us_\theta^2&0\\
  0&-vc_\theta s_\theta&0&vs_\theta^2
 \end{pmatrix}.
 \label{S:noise}
\end{equation}
Here $q$ is the squeezing parameter selected by the canonical covariance
construction.
\end{lemma}

\begin{proof}
We summarize the covariance matrix construction of
Ref.~\cite{IvanSimon2008}.
Local symplectic operations first remove $x$--$p$ correlations while
retaining two independent local squeezing parameters. One then
chooses those scales and a two-mode-squeezing parameter $q$ so that
the position and momentum blocks of $V'-V_q$ are both singular.
Along the resulting one-parameter family, continuity permits the two
null vectors to be chosen with a common angle $\theta$, namely
$(\sin\theta,-\cos\theta)$ in the position block and
$(\sin\theta,\cos\theta)$ in the momentum block.
Positivity makes the complementary blocks rank one and yields the
noise matrix in Eq.~\eqref{S:noise}, with $t=\tan\theta$.
There are two branches; choosing the smaller-squeezing branch gives
$q\leq t$. This is Proposition~1 and condition~(5) of
Ref.~\cite{IvanSimon2008}.
This establishes the canonical covariance decomposition used in the Letter.

It remains to verify the property of the added noise used in the
Letter. Define
\begin{equation}
 X_t=t x_A-x_B,
 \qquad
 P_t=t p_A+p_B .
 \label{S:XP}
\end{equation}
The position and momentum blocks of $N_{t,u,v}$ are proportional to
$(\cos\theta,\sin\theta)(\cos\theta,\sin\theta)^{\mathsf T}$ and
$(\cos\theta,-\sin\theta)(\cos\theta,-\sin\theta)^{\mathsf T}$, respectively.
Since $t=\tan\theta$, $(t,-1)$ is orthogonal to $(\cos\theta,\sin\theta)$ and $(t,1)$ is orthogonal
to $(\cos\theta,-\sin\theta)$, the added noise contributes neither to
$\langle X_t^2\rangle$ nor to $\langle P_t^2\rangle$.
Consequently,
\begin{equation}
 \Tr(\rho_{V'}\widehat D_t)=d_t(q).
 \label{S:witness}
\end{equation}
Moreover, $N_{t,u,v}\geq0$ for $u,v\geq0$, so the covariance
decomposition in Eq.~\eqref{S:canonical} corresponds to adding
classical Gaussian displacement noise to the TMSV covariance.
Together with Eq.~\eqref{S:witness}, this completes the proof.
\end{proof}

\section{Generalized EPR observable and optimal ensemble}
\label{S:eprsec}

This section collects the derivations behind the physical statements
made in the Letter about the observable $\widehat D_t$ and the optimal
Gaussian ensemble.

\subsection{Bogoliubov representation and separability threshold}
For $0<t<1$, the operator $e_t=(b-ta^\dagger)/\sqrt{1-t^2}$ satisfies
$[e_t,e_t^\dagger]=(1-t^2)^{-1}([b,b^\dagger]-t^2[a,a^\dagger])=1$, and
$ta-b^\dagger=-\sqrt{1-t^2}\,e_t^\dagger$. Hence
\begin{equation}
 \widehat D_t=(ta-b^\dagger)^\dagger(ta-b^\dagger)
 =(1-t^2)\,e_te_t^\dagger=(1-t^2)(e_t^\dagger e_t+1),
 \label{S:bogoliubov}
\end{equation}
which is Eq.~\eqref{eq:bogoliubov} of the Letter. On the TMSV, using
$b\lvert\Psi_q\rangle=qa^\dagger\lvert\Psi_q\rangle$,
\begin{equation}
 \langle e_t^\dagger e_t\rangle_{\Psi_q}
 =\frac{(t-q)^2}{(1-t^2)(1-q^2)},
 \label{S:occupation}
\end{equation}
which vanishes if and only if $q=t$. The two EPR quadratures
$X_t=tx_A-x_B$ and $P_t=tp_A+p_B$ satisfy $[X_t,P_t]=i(t^2-1)$, so the
balanced case $t=1$ is the limit in which they commute and no normalizable
Bogoliubov mode is associated with $\widehat D_1$.

For a product state, the local uncertainty relations
$\operatorname{Var}(x_j)+\operatorname{Var}(p_j)\geq1$ give
$\operatorname{Var}(X_t)+\operatorname{Var}(P_t)\geq1+t^2$. Since first
moments only add the nonnegative term $\frac12(\langle X_t\rangle^2+\langle P_t\rangle^2)$,
Eq.~\eqref{eq:Dop} of the Letter gives $\langle\widehat D_t\rangle\geq1$, and
convexity extends the bound to all separable states. This is the threshold
underlying the EPR inseparability criteria of
Refs.~\cite{Duan2000,Marian2018EPR,Marian2021EPR}.

\subsection{Structure of the added noise}
With $t=\tan\theta$, $c_\theta=\cos\theta$, $s_\theta=\sin\theta$, define
$k_x=(c_\theta,0,s_\theta,0)^{\mathsf T}$ and
$k_p=(0,c_\theta,0,-s_\theta)^{\mathsf T}$. The noise matrix of
Eq.~\eqref{eq:noise} of the Letter factorizes as
\begin{equation}
 N_{t,u,v}=\frac{u}{2}k_xk_x^{\mathsf T}
           +\frac{v}{2}k_pk_p^{\mathsf T},
 \label{S:noisefactor}
\end{equation}
so that $N_{t,u,v}\geq0$ and $\operatorname{rank}N_{t,u,v}\leq2$. The EPR
directions $\ell_x=(t,0,-1,0)^{\mathsf T}$ and $\ell_p=(0,t,0,1)^{\mathsf T}$,
which define $X_t=\ell_x^{\mathsf T}R$ and $P_t=\ell_p^{\mathsf T}R$, are
orthogonal to both $k_x$ and $k_p$; hence $N_{t,u,v}\ell_x=N_{t,u,v}\ell_p=0$.
Let $W(\xi)$ denote the local displacement by $\xi=Xk_x+Yk_p$, with $X,Y$
independent centered Gaussian variables of variances $u/2$ and $v/2$. Then
\begin{equation}
 \rho_{V'}=\mathbb E_{X,Y}\left[
 W(\xi)|\Psi_q\rangle\langle\Psi_q|W(\xi)^\dagger\right],
 \label{S:ensemblephysical}
\end{equation}
and $W(\xi)^\dagger X_tW(\xi)=X_t+\ell_x^{\mathsf T}\xi=X_t$, and likewise for
$P_t$. Each component of the ensemble therefore has
$\langle\widehat D_t\rangle=d_t(q)$, as stated in the Letter. Zero variances
are allowed, in which case the ensemble is supported on a line or a point.

\subsection{Classicality and separability of the added noise}
In Ref.~\cite{Marian2008}, the candidate decomposition is described by
adding to the TMSV a classical Gaussian state with covariance
$V_{\rm cl}=I/2+N$, and the optimal $N$ is characterized by requiring
$V_{\rm cl}$ to lie at the boundary of classicality and at the limit of
separability. Both conditions follow from Eq.~\eqref{S:noisefactor}.
First, $\det N_{t,u,v}=0$: the added state is classical, but only
marginally so. Second, $\ell_x$ and $\ell_p$ are eigenvectors of
$V_{\rm cl}$ with eigenvalue $1/2$. After partial transposition
($p_B\to-p_B$), they become $\ell_x$ and $\ell_p'=(0,t,0,-1)^{\mathsf T}$,
which span a plane mapped into itself by $\Omega$ ($\Omega\ell_x=-\ell_p'$,
$\Omega\ell_p'=\ell_x$); on this plane the partially transposed covariance
equals $I/2$. Hence the partial transpose of $V_{\rm cl}$ has a symplectic
eigenvalue equal to $1/2$, i.e., $V_{\rm cl}$ lies exactly on the
separability boundary. In the original frame, the pure covariance of the
ensemble is $\gamma_p=(S_A\oplus S_B)^{-1}V_q(S_A\oplus S_B)^{-\mathsf T}$,
and $V-\gamma_p\geq0$ is the covariance of the displacement noise. We exhibit
one optimal ensemble; we do not claim that it is unique.

\section{Explicit special cases}
\label{S:specialsec}

\subsection{Symmetric states}
For a symmetric two-mode Gaussian state, let $\widetilde\nu_-$ be the
smaller symplectic eigenvalue of the partially transposed covariance
matrix. The result of Ref.~\cite{Giedke2003} reads
\begin{equation}
 E_{\mathrm F}=\mathcal E\!\left(\frac{1-2\widetilde\nu_-}{1+2\widetilde\nu_-}\right)
 =g\!\left(\frac{(1-2\widetilde\nu_-)^2}{8\widetilde\nu_-}\right)
 \quad(\widetilde\nu_-<\tfrac12),
\end{equation}
and zero otherwise. It coincides with the balanced bound
\eqref{eq:balancedbound} of the Letter evaluated at
$\delta=2\widetilde\nu_-$.

\subsection{Two-mode squeezed thermal states}
Consider the standard form with marginal blocks $aI_2,bI_2$ and
correlation block $k\,\mathrm{diag}(1,-1)$, $k\geq0$, which describes
all two-mode squeezed thermal states up to local unitaries. The state
is entangled iff $k^2>(a-\frac12)(b-\frac12)$. In that case, define
\begin{align}
 L&=\sqrt{(a+b)^2-4k^2},\qquad K=2(ab-k^2)+\tfrac12,\nonumber\\
 \vartheta&=\operatorname{artanh}\frac{2k}{a+b}.
\end{align}
Then
\begin{equation}
 r_0=\tfrac12\left[\vartheta-\operatorname{arcosh}(K/L)\right],
 \qquad E_{\mathrm F}=g(\sinh^2r_0),
 \label{S:thermalformula}
\end{equation}
where $r_0$ is the smaller root of $(a+b)\cosh2r-2k\sinh2r=K$, i.e., the
closed formula given in the Letter after Eq.~\eqref{eq:covopt}. To see that
this is the canonical decomposition, put $v_0=\cosh(2r_0)/2$ and
$w_0=\sinh(2r_0)/2$, the entries of $V_q$ with $q=\tanh r_0$. Using
$v_0^2-w_0^2=\frac14$, the defining equation is equivalent to
\begin{equation}
 (a-v_0)(b-v_0)=(k-w_0)^2,
\end{equation}
so the position block of $V-V_q$ is singular; the momentum block has the
opposite off-diagonal sign and is singular as well. These are the two
rank-one blocks of Eq.~\eqref{eq:noise} of the Letter, with $u=v$ and
$t=\sqrt{(b-v_0)/(a-v_0)}$ (exchanging the parties if necessary so that
$t\leq1$), and the smaller root gives $\tanh r_0\leq t$; if $V=V_q$ the
added noise vanishes and one may take $t=1$. Theorem~1 therefore
certifies the Gaussian candidate of Refs.~\cite{Marian2008,Tserkis2017} as
the unrestricted optimum.

\subsection{Channel-generated and extremal families}
Sending one half of a TMSV through a quantum-limited attenuator or
amplifier gives squeezed thermal states at the boundary $q=t$ of the
canonical form; for these, entanglement cost and entanglement of
formation were shown to coincide in Ref.~\cite{Wilde2018}, a result now
extended to all two-mode Gaussian states (Corollary~\ref{cor:superadd}).
For other structured families, including the extremal mixed Gaussian
states of Ref.~\cite{AdessoIlluminati2005}, Eq.~\eqref{eq:main} of the Letter
makes the Gaussian covariance optimization exact for the unrestricted
entanglement of formation and entanglement cost; the Gaussian values
computed or bounded in Ref.~\cite{AdessoIlluminati2005} thereby apply to
$E_{\mathrm F}=E_{\mathrm C}$ itself.

\section{Proof of Theorem~2}
\label{S:additivity}

Throughout this section $A=A_1\cdots A_m$ and $B=B_1\cdots B_m$ consist of
$m$ modes each, with annihilation operators $a_j,b_j$, and
$N_{A_j}=a_j^\dagger a_j$, $N_{B_j}=b_j^\dagger b_j$. We prove Theorem~2 of
the Letter: for every normalized pure state $\lvert\psi\rangle$ of $A{:}B$
with finite mean energy and all $0<q_j<t_j\leq1$,
\begin{equation}
 E(\psi)+\sum_{j=1}^m\lambda_{q_j,t_j}
 \bigl[\langle\widehat{D}^{(j)}\rangle_\psi-d_{t_j}(q_j)\bigr]
 \geq\sum_{j=1}^m\mathcal E(q_j),
 \label{S:thm2}
\end{equation}
where $\widehat{D}^{(j)}$ is the generalized EPR observable of the pair
$(A_j,B_j)$ with gain $t_j$, with the roles of $a_j$ and $b_j$ possibly
exchanged. For $m=1$ this is Theorem~1. The proof below does not use
Sec.~\ref{S:thm1proof}, and therefore provides an independent proof of Theorem~1. Instead
of reducing the problem to Schmidt spectra, it identifies the minimizers of
the left-hand side directly, through their second variations.

\subsection{A minimum principle}
For each pair, let $\alpha_j,\beta_j,\kappa_j>0$ with
$\alpha_j\beta_j\geq\kappa_j^2$, and
\begin{equation}
 T=\sum_jT_j,\qquad
 T_j=\alpha_jN_{A_j}+\beta_jN_{B_j}-\kappa_j(a_jb_j+a_j^\dagger b_j^\dagger).
 \label{S:T}
\end{equation}
Since $\kappa_j\leq\sqrt{\alpha_j\beta_j}\leq(\alpha_j+\beta_j)/2$, the function
$\kappa_j\cosh w-w-(\alpha_j+\beta_j)/2$ is nonpositive at $w=0$, has
derivative $-1$ there, is strictly convex, and diverges. It therefore has a
unique positive root $w_j$, which defines the calibrated squeezing
\begin{equation}
 \kappa_j\cosh w_j-w_j=\frac{\alpha_j+\beta_j}{2},\qquad q_j=e^{-w_j}.
 \label{S:calib}
\end{equation}

\begin{theorem}[Minimum principle]\label{thm:minprinciple}
For every normalized pure state $\lvert\psi\rangle$ of $A{:}B$ with finite mean
energy,
\begin{equation}
 S(\rho_A)+\langle T\rangle_\psi\geq
 \sum_j\left[-\ln(1-q_j^2)-\kappa_jq_j\right],
 \label{S:minprinciple}
\end{equation}
with equality for $\bigotimes_j\lvert\Psi_{q_j}\rangle_{A_jB_j}$.
\end{theorem}

Equality is a direct computation. On $\lvert\Psi_q\rangle$ one has
$\langle N_A\rangle=\langle N_B\rangle=n_q:=q^2/(1-q^2)$,
$\langle ab\rangle=q/(1-q^2)$ and $\mathcal E(q)=-\ln(1-q^2)+2wn_q$; by
Eq.~\eqref{S:calib}, $(\alpha+\beta)n_q=[\kappa(q+q^{-1})-2w]n_q$, and the
terms combine to $-\ln(1-q^2)-\kappa q$.

\emph{Theorem~2 from Theorem~\ref{thm:minprinciple}.} By Eq.~\eqref{eq:Dop}
of the Letter, $\lambda\widehat D_t=\lambda(t^2N_A+N_B+1)-\lambda t(ab+a^\dagger b^\dagger)$.
Thus $\lambda_j\widehat{D}^{(j)}=T_j+\lambda_j$ with
$(\alpha_j,\beta_j,\kappa_j)=\lambda_j(t_j^2,1,t_j)$, or $\lambda_j(1,t_j^2,t_j)$
if the roles of $a_j$ and $b_j$ are exchanged; in both cases
$\alpha_j\beta_j=\kappa_j^2$. Since
$t\cosh w-(1+t^2)/2=(t-q)(1-tq)/(2q)$ for $q=e^{-w}$, the calibration
\eqref{S:calib} is equivalent to $\lambda_j=\lambda_{q_j,t_j}$. A short
computation gives
$\mathcal E(q)+\lambda_{q,t}d_t(q)=-\ln(1-q^2)-\lambda_{q,t}tq+\lambda_{q,t}$,
so Eq.~\eqref{S:minprinciple} is Eq.~\eqref{S:thm2}.

\subsection{Stationarity and second variations}
Represent $\lvert\psi\rangle$ by its coefficient operator $C$ as in
Eq.~\eqref{S:Cstate}, now in the product Fock bases of $A$ and $B$, so that
$\rho_A=CC^\dagger$, and write $TC$ for the coefficient operator of
$T\lvert\psi\rangle$. The objective is
\begin{equation}
 \mathcal F(C)=-\Tr[CC^\dagger\ln(CC^\dagger)]+\langle C,TC\rangle,
 \qquad \langle X,Y\rangle=\Tr X^\dagger Y.
\end{equation}
In Schmidt bases, $\lvert\psi\rangle=\sum_lc_l\lvert u_l\rangle\lvert v_l\rangle$ and
$C=\operatorname{diag}(c_l)$, with $p_l=c_l^2$. The two marginal kernels
may have different dimensions. We therefore enlarge each local space by an
auxiliary direct summand, on which all operators act as zero, and pair the
enlarged kernels; states, variations and expectation values are unchanged.
This enlargement is only a device for the matrix notation: minimizations,
admissible variations and compactness arguments below refer to the
physical Fock spaces, which the test operators preserve.
Let $\mathcal G(C)=\ln(CC^\dagger)\,C$, with $0\ln0=0$. Since
$\delta\mathcal F=2\operatorname{Re}\langle\delta C,TC-\mathcal G(C)-C\rangle$,
the stationarity condition on the unit sphere reads
\begin{equation}
 TC-\mathcal G(C)=\mu C,\qquad \mu=\mathcal F(C).
 \label{S:euler}
\end{equation}

By the Daleckii--Krein formula, the real derivative of $\mathcal G$ at
$C=\operatorname{diag}(c_l)$ is
\begin{equation}
 [D\mathcal G(Y)]_{lr}=F_{lr}Y_{lr}+\sqrt{p_lp_r}\,L_{lr}\overline{Y_{rl}},
 \label{S:DG}
\end{equation}
where $L_{lr}=(\ln p_l-\ln p_r)/(p_l-p_r)$ and
\begin{equation}
 F_{lr}=\frac{p_l\ln p_l-p_r\ln p_r}{p_l-p_r}
 =\ln p_l+p_rL_{lr}=\ln p_r+p_lL_{lr},
\end{equation}
with the continuous values $F_{ll}=1+\ln p_l$, and $F_{lr}=\ln p_l$,
$\sqrt{p_lp_r}L_{lr}=0$ if $p_r=0<p_l$. For a normalized curve through a
stationary point with tangent $Y$, $\langle C,Y\rangle=0$, the second
derivative of $\mathcal F$ is
$2\operatorname{Re}\langle Y,(T-\mu-D\mathcal G)Y\rangle$. The last term in
Eq.~\eqref{S:DG} changes sign under $Y\to iY$. Averaging the second variations
along $Y$ and $iY$ therefore gives the complex-linear form
\begin{equation}
 Q(Y)=\langle Y,H_0Y\rangle,\qquad H_0=T-\mu-\mathsf F,
 \label{S:Q}
\end{equation}
where $\mathsf F$ multiplies the $(l,r)$ entry by $F_{lr}$. At a local
minimum, $Q(Y)\geq0$ whenever $\langle C,Y\rangle=0$. Because
$F_{ll}=1+\ln p_l$, Eq.~\eqref{S:euler} gives $H_0C=-C$, and evaluating $Q$
on $Y-\langle C,Y\rangle C$ yields, for arbitrary $Y$,
\begin{equation}
 Q(Y)\geq-\lvert\langle C,Y\rangle\rvert^2 .
 \label{S:stability}
\end{equation}
Domains and the differentiability of the entropy are established in
Lemma~\ref{lem:filters} and in the regularization argument below, for the regularized problem
used there; here we derive the algebraic consequences of
Eqs.~\eqref{S:euler}--\eqref{S:stability}.

\subsection{Two test directions}
For each pair, consider the operators
\begin{equation}
 O_{1,j}=a_j-q_jb_j^\dagger,\qquad O_{2,j}=q_ja_j^\dagger-b_j,
 \label{S:tests}
\end{equation}
which annihilate $\bigotimes_j\lvert\Psi_{q_j}\rangle$, since
$a\lvert\Psi_q\rangle=qb^\dagger\lvert\Psi_q\rangle$ and
$b\lvert\Psi_q\rangle=qa^\dagger\lvert\Psi_q\rangle$. Let
$A_{j,lr}=\langle u_l\rvert a_j\lvert u_r\rangle$ and
$B_{j,lr}=\langle v_l\rvert b_j\lvert v_r\rangle$. The coefficient operators of
$a_j\lvert\psi\rangle$ and $b_j^\dagger\lvert\psi\rangle$ have entries
$A_{j,lr}c_r$ and $c_l\overline{B_{j,lr}}$; neither has kernel-to-kernel
entries. In this notation $\langle N_{A_j}\rangle=\sum_{lr}p_r|A_{j,lr}|^2$
and $\langle N_{A_j}\rangle+1=\sum_{lr}p_l|A_{j,lr}|^2$, so that
\begin{equation}
 \sum_{lr}(p_l-p_r)|A_{j,lr}|^2=\sum_{lr}(p_l-p_r)|B_{j,lr}|^2=1,
 \label{S:wccr}
\end{equation}
and $k_j:=\operatorname{Re}\langle a_jb_j\rangle=\sum_{lr}c_lc_r\operatorname{Re}(A_{j,lr}B_{j,lr})$.

For $O=O_{1,j}$, Eq.~\eqref{S:euler} gives
$(T-\mu)O\lvert\psi\rangle=O\,\mathcal G(C)+[T,O]\lvert\psi\rangle$, with
$[T,a_j]=-\alpha_ja_j+\kappa_jb_j^\dagger$ and
$[T,b_j^\dagger]=\beta_jb_j^\dagger-\kappa_ja_j$, and similarly for $O_{2,j}$.
In the entropy terms,
$\ln p_r-F_{lr}=-p_lL_{lr}$ and $\ln p_l-F_{lr}=-p_rL_{lr}$. Dropping the
index $j$ on $\alpha,\beta,\kappa,q$, one finds
\begin{align}
 Q(O_{1,j}\psi)={}&(q\kappa-\alpha)n_A+(q^2\beta-q\kappa)(n_B+1)\nonumber\\
 &+[\kappa(1-q^2)+q(\alpha-\beta)]k_j\nonumber\\
 &-\mathsf A_j-q^2\mathsf B_j+q\mathsf C_j,\nonumber\\
 Q(O_{2,j}\psi)={}&(q^2\alpha-q\kappa)(n_A+1)+(q\kappa-\beta)n_B\nonumber\\
 &+[\kappa(1-q^2)-q(\alpha-\beta)]k_j\nonumber\\
 &-q^2\mathsf A_j-\mathsf B_j+q\mathsf C_j,
 \label{S:Q12}
\end{align}
where $n_A=\langle N_{A_j}\rangle$, $n_B=\langle N_{B_j}\rangle$, and
\begin{align}
 \mathsf A_j&=\sum_{lr}p_lp_rL_{lr}|A_{j,lr}|^2,\qquad
 \mathsf B_j=\sum_{lr}p_lp_rL_{lr}|B_{j,lr}|^2,\nonumber\\
 \mathsf C_j&=\sum_{lr}\sqrt{p_lp_r}\,(p_l+p_r)L_{lr}\operatorname{Re}(A_{j,lr}B_{j,lr}).
\end{align}
Now set $x_{lr}=\frac12\ln(p_l/p_r)$, so that $c_lc_re^{x_{lr}}=p_l$,
$c_lc_re^{-x_{lr}}=p_r$, $p_lp_rL_{lr}=c_lc_rf(x_{lr})$ and
$\sqrt{p_lp_r}(p_l+p_r)L_{lr}=2c_lc_rb(x_{lr})$, with
$f(x)=x/\sinh x$ and $b(x)=x\coth x$. Adding the two lines of
Eq.~\eqref{S:Q12}, and using Eq.~\eqref{S:wccr} to write the constants
$q^2\beta-q\kappa$ and $q^2\alpha-q\kappa$ as weighted sums over $(l,r)$, we
obtain the exact identity
\begin{equation}
 Q(O_{1,j}\psi)+Q(O_{2,j}\psi)=-\sum_{lr}c_lc_r\,
 z_{j,lr}^\dagger M_j(x_{lr})z_{j,lr},
 \label{S:mainid}
\end{equation}
with $z_{j,lr}=(A_{j,lr},\overline{B_{j,lr}})^{\mathsf T}$ and
\begin{align}
 M_{11}(x)&=q(\kappa-q\alpha)e^x+(\alpha-q\kappa)e^{-x}+(1+q^2)f(x),\nonumber\\
 M_{22}(x)&=q(\kappa-q\beta)e^x+(\beta-q\kappa)e^{-x}+(1+q^2)f(x),\nonumber\\
 M_{12}(x)&=M_{21}(x)=-\kappa(1-q^2)-2q\,b(x).
 \label{S:M}
\end{align}
Terms with exactly one vanishing eigenvalue are understood as the
continuous limits of the coefficients, e.g., $c_lc_re^{x_{lr}}=p_l$; these
terms are retained. In the same notation, the norms of the test vectors are
\begin{equation}
 \|O_{1,j}\psi\|^2+\|O_{2,j}\psi\|^2=\sum_{lr}c_lc_r\,
 z_{j,lr}^\dagger K_j(x_{lr})z_{j,lr},
 \label{S:defects}
\end{equation}
\begin{equation}
 K_j(x)=\begin{pmatrix}e^{-x}+q^2e^x&-2q\\-2q&e^{-x}+q^2e^x\end{pmatrix}.
\end{equation}

\subsection{The matrix inequality}
\begin{lemma}\label{lem:matrix}
Let $\alpha,\beta,\kappa>0$ with $\alpha\beta\geq\kappa^2$, and $q=e^{-w}$ as in
Eq.~\eqref{S:calib}. Then $M(x)\geq0$ for all real $x$, with
$\ker M(x)=\{0\}$ for $x\neq w$ and $\ker M(w)=\operatorname{span}\{(1,1)\}$.
Moreover, there is $\gamma>0$, depending only on $\alpha,\beta,\kappa$, such that
$M(x)\geq\gamma K(x)$ for all $x$. For several pairs we write $\gamma=\min_j\gamma_j$.
\end{lemma}

\begin{proof}
Write $\mathcal C_w=\cosh w$, $\mathcal S_w=\sinh w$, $\delta=\alpha-\beta$ and
$y=x-w$. Using Eq.~\eqref{S:calib},
\begin{equation}
 \frac{M(x)}{2q}=
 \begin{pmatrix}
 m_+(x)&-\kappa\mathcal S_w-b(x)\\
 -\kappa\mathcal S_w-b(x)&m_-(x)
 \end{pmatrix},
 \label{S:Mtilde}
\end{equation}
with $m_\pm(x)=\kappa\mathcal S_w\cosh y+(w\mp\frac\delta2)\sinh y+\mathcal C_wf(x)$.
Since $\alpha+\beta\geq2\kappa$, Eq.~\eqref{S:calib} gives
$\kappa\geq\kappa_0:=w/(\mathcal C_w-1)$, hence $\kappa\mathcal S_w\geq w\coth(w/2)>w$,
and the trace of \eqref{S:Mtilde} is positive. Let $\Delta_0(x)$ be its
determinant at $\kappa=\kappa_0$, $\delta=0$. Expanding,
\begin{align}
 \det\frac{M(x)}{2q}={}&\Delta_0(x)
 +2(\kappa-\kappa_0)\sinh(x-w)\bigl[w\sinh x-\mathcal S_w^2f(x)\bigr]\nonumber\\
 &+(\alpha\beta-\kappa^2)\sinh^2(x-w).
 \label{S:detid}
\end{align}
The last term is nonnegative. So is the middle one: for $x\leq0$ both factors
are negative, while for $x>0$ the bracket has the sign of $x-w$, because
$\sinh^2x/x$ is increasing. At $\kappa=\kappa_0$, $\delta=0$, the diagonal
entries of \eqref{S:Mtilde} coincide, the eigenvalue along $(1,-1)$ is
positive, and the eigenvalue along $(1,1)$ equals
\begin{align}
 &\frac{2w}{\sinh(w/2)}\sinh\frac{x-w}{2}\,I(x),\nonumber\\
 &I(x)=\sinh\frac x2-\frac{\sinh(w/2)}{w}\,\frac{x\sinh\frac{x+w}2}{\sinh x}.
\end{align}
For $x>0$, the function
$\mathcal A(x)=\sinh x\sinh(x/2)/[x\sinh((x+w)/2)]$ has logarithmic derivative
$\coth x-x^{-1}+\frac12[\coth\frac x2-\coth\frac{x+w}2]>0$ and
$\mathcal A(w)=\sinh(w/2)/w$, so $I(x)$ has the sign of $x-w$. For
$-w\leq x\leq0$, $I(x)<0$ directly. For $x=-z<-w$,
$I(-z)=-\sinh\frac z2+\frac{\sinh(w/2)}{w}\frac{z\sinh((z-w)/2)}{\sinh z}<0$,
since $z/\sinh z<w/\sinh w$ and $\sinh\frac{z-w}2<\sinh\frac z2$. Hence
$\Delta_0(x)>0$ for $x\neq w$, with a zero of second order at $x=w$. With
Eq.~\eqref{S:detid}, this proves the statements on positivity and on the kernel.

For the comparison with $K$, which is positive definite except for the same
kernel at $x=w$, three regions are relevant. Near $x=w$, in the basis
$(1,1)/\sqrt2$, $(1,-1)/\sqrt2$ and after conjugation by
$\operatorname{diag}[(x-w)^{-1},1]$, $M$ and $K$ tend to
\begin{equation}
 2q\begin{pmatrix}\frac12(\kappa\mathcal S_w-2+w\coth w)&-\frac\delta2\\
 -\frac\delta2&2(\kappa\mathcal S_w+w\coth w)\end{pmatrix}
\end{equation}
and $\operatorname{diag}(q,4q)$, respectively; the first matrix is positive definite by the second-order zero of the
determinant. As $x\to+\infty$ the generalized eigenvalues of $M$ relative
to $K$ tend to $(\kappa-q\alpha)/q$ and $(\kappa-q\beta)/q$, and as $x\to-\infty$
to $\alpha-q\kappa$ and $\beta-q\kappa$. These are positive: $\alpha,\beta$ lie
between the roots of $X^2-(\alpha+\beta)X+\kappa^2$, whose sum
$\alpha+\beta=\kappa(q+q^{-1})-2w$ is smaller than $\kappa(q+q^{-1})$, the sum of
the roots $q\kappa$ and $\kappa/q$ of $X^2-\kappa(q+q^{-1})X+\kappa^2$; hence
$q\kappa<\alpha,\beta<\kappa/q$. Compactness of the remaining interval completes
the proof.
\end{proof}

\emph{Rigidity.} If $\lvert\psi\rangle$ is a local minimizer at which
Eqs.~\eqref{S:euler}--\eqref{S:stability} hold for the test vectors, and
$\langle O_{i,j}\rangle_\psi=0$, then
Eqs.~\eqref{S:mainid}, \eqref{S:defects} and Lemma~\ref{lem:matrix} give
$0\leq\sum_j[Q(O_{1,j}\psi)+Q(O_{2,j}\psi)]\leq-\gamma\sum_{i,j}\|O_{i,j}\psi\|^2$,
so all test vectors vanish. The operators
$f_j=O_{1,j}/\sqrt{1-q_j^2}$ and $e_j=-O_{2,j}/\sqrt{1-q_j^2}$ are
independent canonical annihilators, whose joint vacuum is
$\bigotimes_j\lvert\Psi_{q_j}\rangle$. More quantitatively, since their total
number operator has gap one above this vacuum,
\begin{equation}
 \sum_{i,j}\|O_{i,j}\psi\|^2\geq(1-q_{\max}^2)
 \Bigl(1-\bigl\lvert\langle\textstyle\bigotimes_j\Psi_{q_j}\vert\psi\rangle\bigr\rvert^2\Bigr).
 \label{S:gap}
\end{equation}
The minimizer is thus the product of two-mode squeezed vacua, without any
prior assumption on its Schmidt bases. It remains to show that minimizers
exist and that the formal steps are legitimate.

\subsection{Entropy variations under local filters}
\label{S:filters}
\begin{lemma}\label{lem:filters}
Let $C$ be Hilbert--Schmidt with $\sum_lp_l(1+\ln^2p_l)<\infty$, let $X$ and
$Y$ be bounded operators on $A$ and $B$, and let
$\Phi(s,t)=\Tr h(C_{s,t}C_{s,t}^\dagger)$, with $h(z)=z\ln z$ and
$C_{s,t}=(I+sX)C(I+tY)^{\mathsf T}$. Then $\Phi$ is twice continuously
differentiable near $(0,0)$, and its second derivatives are the limits of
those obtained by replacing $C$ with its Schmidt truncations $C_n$. In
particular, the second variations of the entropy along local-filter curves
are obtained by differentiating $\mathcal G$ as in Eq.~\eqref{S:DG}; at a
normalized stationary point, they yield the constrained second variation
used in Eqs.~\eqref{S:Q}--\eqref{S:stability} along the tangent
$XC+CY^{\mathsf T}$.
\end{lemma}

\begin{proof}
For small $|s|,|t|$ the filters are invertible, so $C_{s,t}$ and $C$ have the
same rank and comparable singular values, and no new eigenvalue emerges
from the kernel. For fixed $t$, the nonzero spectrum of
$C_{n;s,t}C_{n;s,t}^\dagger$ equals that of $\Gamma(s)=R^\dagger P(s)R$, with
$R=C_n(I+tY)^{\mathsf T}$ and $P(s)=(I+sX)^\dagger(I+sX)$. Thus $\Gamma$ is quadratic
in $s$, and $\Gamma'=\Gamma^{1/2}\Xi_1\Gamma^{1/2}$, $\Gamma''=\Gamma^{1/2}\Xi_2\Gamma^{1/2}$ with $\Xi_i$
bounded uniformly in $n$, since $P(s)$ is uniformly invertible. The
resolvent representation
$D^2\Tr h(\Gamma)[V_1,V_2]=\int_0^\infty\Tr[V_1(\Gamma+r)^{-1}V_2(\Gamma+r)^{-1}]dr$ and its
higher-order analogues, together with H\"older's inequality and
$\int_0^\infty z^k(z+r)^{-k}dr=z/(k-1)$, give
$|D^k\Tr h(\Gamma)[V_1,\ldots,V_k]|\leq c_k\Tr\Gamma\prod_i\|\Xi_i\|$ for $k\geq2$. Hence
$\partial_s^4\Phi_n$ is bounded uniformly in $n$. For $\partial_t^4\Phi_n$ we
use instead $(I+sX)C_nP'(t)C_n^\dagger(I+sX)^\dagger$, with
$P'(t)=(I+tY)^{\mathsf T}(I+tY)^{\mathsf T\dagger}$, which has the same nonzero
spectrum and is quadratic in $t$; the same bound follows. No lower bound on
the Schmidt coefficients is needed. The operator $\partial_s^4+\partial_t^4$ is elliptic,
so the interior estimate bounds $\Phi_n$ in $W^{4,2}$ on a smaller rectangle,
uniformly in $n$, and the compact embedding $W^{4,2}\subset C^{2,\alpha}$, $0<\alpha<1$, in two
dimensions makes $\{\Phi_n\}$ precompact in $C^2$. Since $\Phi_n\to\Phi$
uniformly (the tails are dominated by $\sum_lp_l(1+|\ln p_l|)$), $\Phi$ is
$C^2$ and the second derivatives converge. For finite rank, the Hessian is
given by Eq.~\eqref{S:DG}, and the spectral sums converge absolutely by
$|F_{lr}|\leq1+|\ln\max(p_l,p_r)|$ and $0\leq\sqrt{p_lp_r}L_{lr}\leq1$, since
every entry of $XC+CY^{\mathsf T}$ carries a factor $c_l$ or $c_r$.
\end{proof}

\subsection{Regularization and proof of Theorem~\ref{thm:minprinciple}}
\label{S:regularization}

We apply the stability conditions to global minimizers of a
regularized functional and then show that their values converge
to the unrestricted infimum.

Assume first $\alpha_j\beta_j>\kappa_j^2$ for all $j$. Then $T\geq\epsilon N-\tau$
for some $\epsilon>0$, where $N=\sum_j(N_{A_j}+N_{B_j})$. Let $W=1+N$, fix an integer
$r\geq2$, and, for $\zeta>0$, consider
\begin{equation}
 \mathcal F_\zeta(C)=\mathcal F(C)+\zeta\langle C,W^rC\rangle,
\end{equation}
with infimum $v_\zeta$. Let $v_0$ be the infimum of $\mathcal F$ over
finite-energy states.

\emph{Existence.} Minimizing sequences of $\mathcal F_\zeta$ have bounded
$\langle W^r\rangle$; since $W$ has compact resolvent, they are precompact in
norm. The entropy and the positive forms $\langle T+\tau\rangle$,
$\langle W^r\rangle$ are lower semicontinuous, so a minimizer $C_\zeta$ exists.
Finite Fock vectors approximate any finite-energy state with convergent
energy and, by energy-constrained continuity \cite{Winter2016}, convergent
entropy. Hence $v_\zeta\downarrow v_0$, and
$0\leq\zeta\langle W^r\rangle_{C_\zeta}\leq v_\zeta-v_0\to0$.

\emph{Regularity.} For a finite-energy $m$-mode marginal with decreasing
eigenvalues $p_i$, passive rearrangement gives $\sum_ip_i\epsilon_i\leq\langle N_A\rangle$,
with oscillator energies $\epsilon_i\asymp i^{1/m}$; hence
$p_i\leq c\,i^{-1-1/m}$, and $\Tr\rho_A|\ln\rho_A|^s<\infty$ for all $s>0$. In
particular $\mathcal G(C_\zeta)$ is Hilbert--Schmidt. The first variation of
the entropy along a finite-rank perturbation equals
$-2\operatorname{Re}\langle Y,\mathcal G(C)+C\rangle$, since the function $s^2\ln s^2$,
applied to the self-adjoint dilation of $C$, is $C^1$ with vanishing
derivative at $s=0$. Variations along finite Fock vectors therefore give
Eq.~\eqref{S:euler} with $T$ replaced by $T+\zeta W^r$, first weakly. As $T$ is
infinitesimally bounded relative to $W^r$, $T+\zeta W^r$ is self-adjoint on
the domain of $W^r$ with the finite Fock vectors as a core; the equation thus
holds in the operator sense, and $\langle W^{2r}\rangle_{C_\zeta}<\infty$. For any
ladder operator $O$, the identity
$(T+\zeta W^r)O\psi_\zeta=O[\mathcal G(C_\zeta)+\mu_\zeta C_\zeta]+[T+\zeta W^r,O]\psi_\zeta$
then has a right-hand side of finite norm, by the bound
$\Tr\rho|\ln\rho|^2N\leq(\Tr\rho|\ln\rho|^4)^{1/2}(\Tr\rho N^2)^{1/2}$, which
justifies the manipulations leading to Eq.~\eqref{S:Q12}. The spectral sums
there converge absolutely: $c_lc_rb(x_{lr})\leq\frac12(p_l+p_r)(1+\frac12|\ln p_l|+\frac12|\ln p_r|)$,
so that, by Cauchy--Schwarz, $\mathsf C_j$ is bounded by moments
$\Tr\rho_X|\ln\rho_X|(N_X+1)$, $X=A,B$, which are finite by the same bound.
Since $M_j(x)\geq0$, the terms of Eq.~\eqref{S:mainid} are nonnegative, and
the regrouping leading to it is legitimate.

\emph{Second variations.} The cutoffs $a_j\openone_{N_{A_j}\leq L}$ and
$b_j\openone_{N_{B_j}\leq L}$ are bounded and preserve the energy domain.
Lemma~\ref{lem:filters} applied to them, and minimality of $C_\zeta$, give
Eq.~\eqref{S:stability} for the truncated test vectors. As $L\to\infty$, these
converge in the form norm of $T+\zeta W^r$, and the entropy forms converge by
dominated convergence, using the logarithmic moments above. Thus
Eq.~\eqref{S:stability} holds for the test vectors \eqref{S:tests} at $C_\zeta$.

\emph{Penalty and centering errors.} With the penalty,
Eq.~\eqref{S:mainid} acquires the additional term
\begin{equation}
 E_\zeta=\zeta\operatorname{Re}\sum_{i,j}
 \langle O_{i,j}\psi_\zeta,[W^r,O_{i,j}]\psi_\zeta\rangle .
\end{equation}
Since $W^ra=a(W-1)^r$ and $W^ra^\dagger=a^\dagger(W+1)^r$, each commutator is a
ladder operator times a polynomial of degree $r-1$ in $N$, and
$|E_\zeta|\leq c_r\,\zeta\langle W^r\rangle_{C_\zeta}\to0$. Local displacements leave
the entropy invariant, so stationarity under them gives
\begin{equation}
 \begin{pmatrix}\alpha_j&-\kappa_j\\-\kappa_j&\beta_j\end{pmatrix}
 \begin{pmatrix}\langle a_j\rangle\\ \overline{\langle b_j\rangle}\end{pmatrix}
 =\zeta\begin{pmatrix}\langle[W^r,a_j]\rangle\\ \overline{\langle[W^r,b_j]\rangle}\end{pmatrix},
\end{equation}
where, by the same polynomial commutator estimate,
$\zeta|\langle[W^r,a_j]\rangle|\leq c_r\zeta\langle W^r\rangle\to0$, and likewise for
$b_j$. Invertibility of the matrix then gives $\langle a_j\rangle,\langle b_j\rangle\to0$, so
that the right-hand side of Eq.~\eqref{S:stability} also vanishes in the limit.

\emph{Conclusion.} Combining these estimates with Eq.~\eqref{S:mainid}, the
penalized analogue of $H_0C=-C$, and Lemma~\ref{lem:matrix},
\begin{equation}
 \gamma\sum_{i,j}\|O_{i,j}\psi_\zeta\|^2\leq|E_\zeta|
 +\sum_{i,j}|\langle O_{i,j}\rangle_{\psi_\zeta}|^2\longrightarrow0 .
\end{equation}
By Eq.~\eqref{S:gap}, $\psi_\zeta\to\bigotimes_j\lvert\Psi_{q_j}\rangle$ in norm, up to
a phase. By the definition of $v_0$ and lower semicontinuity of $\mathcal F$,
\begin{equation}
 v_0
 \leq \mathcal F\!\left(\bigotimes_j\Psi_{q_j}\right)
 \leq \liminf_{\zeta\downarrow0}\mathcal F(\psi_\zeta)
 \leq \lim_{\zeta\downarrow0}v_\zeta
 =v_0.
\end{equation}
Hence the product of squeezed vacua attains the unrestricted
infimum and is a global minimizer. This proves
Theorem~\ref{thm:minprinciple} for $\alpha_j\beta_j>\kappa_j^2$.

If $\alpha_j\beta_j=\kappa_j^2$ for some $j$, as for Theorem~2, replace
$\alpha_j,\beta_j$ by $\alpha_j+\varepsilon,\beta_j+\varepsilon$. The calibrated
$q_j(\varepsilon)$ tends to $q_j$ as $\varepsilon\downarrow0$, and for each fixed
finite-energy state both sides of Eq.~\eqref{S:minprinciple} converge. The
limits are taken in this order, first $\zeta\downarrow0$ at fixed $\varepsilon$,
so no estimate uniform in $\varepsilon$ is needed. This completes the proof of
Theorem~\ref{thm:minprinciple}, and hence of Theorem~2.

\section{Consequences of Theorem~2}
\label{S:consequences}

\subsection{Superadditivity bound, additivity and entanglement cost}
\begin{corollary}\label{cor:superadd}
Let $\omega$ be a finite-energy state of a bipartite system containing $m$
disjoint mode pairs $(A_j,B_j)$, with $A_j$ on Alice's side and $B_j$ on Bob's
side, and let $V_j$ be the covariance matrix of the pair $(A_j,B_j)$. Then
\begin{equation}
 E_{\mathrm F}(\omega)\geq\sum_jE_{\mathrm F}^{\mathrm G}(V_j).
 \label{S:superadd}
\end{equation}
For every collection of two-mode Gaussian states $\sigma_j$, with covariance matrices $V_{\sigma_j}$,
\begin{equation}
 E_{\mathrm F}\Bigl(\bigotimes_j\sigma_j\Bigr)=\sum_jE_{\mathrm F}(\sigma_j),
 \qquad
 E_{\mathrm C}(\sigma_j)=E_{\mathrm F}(\sigma_j)=E_{\mathrm F}^{\mathrm G}(V_{\sigma_j}).
 \label{S:additive}
\end{equation}
\end{corollary}

\begin{proof}
Discarding all other modes, and the pairs with separable $V_j$, is a local
operation and does not increase $E_{\mathrm F}$. For each remaining pair,
local displacements and local symplectic transformations on $A_j$ and on
$B_j$, which are local unitaries for the bipartition, bring $V_j$ to the
canonical form $V_{q_j}+N_{t_j,u_j,v_j}$ of Lemma~\ref{lem:canonical}, possibly
with the roles of $A_j$ and $B_j$ exchanged; then
$\Tr(\omega\widehat{D}^{(j)})=d_{t_j}(q_j)$ by Eq.~\eqref{S:witness}. Applying
Eq.~\eqref{S:thm2} to every component of a pure-state decomposition of
$\omega$ (almost every component has finite energy) and averaging gives
$\sum_kp_kE(\psi_k)\geq\sum_j\mathcal E(q_j)=\sum_jE_{\mathrm F}^{\mathrm G}(V_j)$; pairs with
$q_j=t_j$ are handled by the limit $q_j'\uparrow t_j$ as in Sec.~\ref{S:thm1proof}. For
$\omega=\bigotimes_j\sigma_j$, the product of the optimal displacement
ensembles of Sec.~\ref{S:eprsec} attains the bound, and $E_{\mathrm F}^{\mathrm G}(V_{\sigma_j})=E_{\mathrm F}(\sigma_j)$ by
Eq.~\eqref{eq:main} of the Letter. Taking $\sigma_j=\sigma$ for all $j$ gives
$E_{\mathrm F}(\sigma^{\otimes m})=mE_{\mathrm F}(\sigma)$. Gaussian states have finite
local entropy, so the entanglement cost equals the regularized
entanglement of formation \cite{Hayden2001,Yamasaki2025}, which proves the
last equality.
\end{proof}

Several features of Eq.~\eqref{S:superadd} are worth noting.
(i) It allows arbitrary correlations, Gaussian or not, between the pairs.
When the pair marginals $\omega_j$ are Gaussian, it reads
$E_{\mathrm F}(\omega)\geq\sum_jE_{\mathrm F}(\omega_j)$, i.e., the entanglement of formation
is strongly superadditive on this class, although strong superadditivity
fails in general \cite{Shor2004,Hastings2009}.
(ii) Applied to $\rho^{\otimes m}$ for an arbitrary finite-energy two-mode
state $\rho$, it bounds the entanglement cost of $\rho$ from its covariance
matrix alone; this is proven in Sec.~\ref{S:nongaussian}, together with the
other covariance-based bounds of the Letter.
(iii) The Gaussian entanglement of formation is additive as well, since
$E_{\mathrm F}\leq E_{\mathrm F}^{\mathrm G}$ on products of Gaussian states; this
includes the symmetric case of Ref.~\cite{Wolf2004}. Equation~\eqref{S:additive}
contains the channel-generated states of Ref.~\cite{Wilde2018} as the special
case $q=t$, $u=v$ of the canonical form.
(iv) By Eq.~\eqref{eq:bisym} of the Letter, bisymmetric Gaussian states, and
more generally multimode Gaussian states that are locally Gaussian-unitarily equivalent to
products of two-mode states and uncorrelated local modes, also satisfy
$E_{\mathrm C}=E_{\mathrm F}=E_{\mathrm F}^{\mathrm G}$, the latter given by the sum of the
two-mode values.

\subsection{Entanglement cost of bosonic Gaussian channels}
Following Ref.~\cite{Wilde2018}, the entanglement cost $E_{\mathrm C}(\mathcal N)$ of a
channel $\mathcal N$ is the minimum entanglement per use required, together with free LOCC, to simulate $n$ uses of $\mathcal N$ with vanishing error against arbitrary adaptive sequential tests. We express this cost in nats per use: consumption of a maximally entangled state of Schmidt rank $M_n$ has rate $\ln M_n/n$, and division by $\ln2$ gives the rate in ebits. The parallel cost
$E_{\mathrm C}^{(\mathrm p)}(\mathcal N)\leq E_{\mathrm C}(\mathcal N)$ restricts the
discriminator to parallel tests. We consider the single-mode thermal-loss
channel $\mathcal L_{\eta,N_B}$, $b\mapsto\sqrt\eta\,a+\sqrt{1-\eta}\,e$, the
thermal amplifier $\mathcal A_{G,N_B}$, $b\mapsto\sqrt G\,a+\sqrt{G-1}\,e^\dagger$,
with the environment mode $e$ in a thermal state with mean photon number
$N_B$, and the additive-noise channel $\mathcal T_\xi$, which adds Gaussian
noise of variance $\xi$ to both quadratures. We take $0<\eta<1$, $G>1$, $N_B\geq0$ and $\xi>0$; identity and replacement channels are treated by their limiting values. For each finite $N_S\geq0$, define
\begin{align}
 q(N_S)&=\sqrt{\frac{N_S}{N_S+1}},\qquad
 \psi_{N_S}=|\Psi_{q(N_S)}\rangle\langle\Psi_{q(N_S)}|,\nonumber\\
 \sigma_{\mathcal N}(N_S)&=(\mathrm{id}\otimes\mathcal N)(\psi_{N_S}).
\end{align}
where $|\Psi_q\rangle$ is the TMSV of the Letter, so $\psi_{N_S}$ has mean photon number $N_S$ per mode. Every resource state in the argument below is a normalizable finite-energy state; infinite squeezing enters only through a limit of entanglement values.

\begin{corollary}\label{cor:channels}
For $\mathcal N\in\{\mathcal L_{\eta,N_B},\mathcal A_{G,N_B},\mathcal T_\xi\}$,
\begin{align}
 E_{\mathrm C}(\mathcal N)&=E_{\mathrm C}^{(\mathrm p)}(\mathcal N)
 =\sup_{N_S\geq0}E_{\mathrm F}(\sigma_{\mathcal N}(N_S))\nonumber\\
 &=g\!\left(\frac{(1-\delta_{\mathcal N})^2}{4\delta_{\mathcal N}}\right)
 \label{S:channelcost}
\end{align}
if $0<\delta_{\mathcal N}<1$. If $\delta_{\mathcal N}\geq1$, all three quantities preceding the last equality are zero. The identity-channel endpoint $\delta_{\mathcal N}=0$ has infinite cost. Here
\begin{equation}
 \delta_{\mathcal N}=\frac{|1-\sqrt\tau|}{1+\sqrt\tau}
 \bigl(\sqrt{N_B+1}+\sqrt{N_B}\bigr)^2
 \label{S:deltaN}
\end{equation}
with $\tau=\eta$ for $\mathcal L_{\eta,N_B}$ and $\tau=G$ for $\mathcal A_{G,N_B}$, and
$\delta_{\mathcal N}=\xi$ for $\mathcal T_\xi$.
\end{corollary}

\begin{proof}
\emph{Lower bound.} Alice can prepare $n$ copies of $\psi_{N_S}$ locally and
send one arm of each through a parallel simulation of $\mathcal N^{\otimes n}$
with error $\varepsilon$. This is an LOCC protocol that consumes the
simulation's ebits and outputs a state within $\varepsilon$ of
$\sigma_{\mathcal N}(N_S)^{\otimes n}$ in trace norm. Hence
$E_{\mathrm C}(\sigma_{\mathcal N}(N_S))\leq E_{\mathrm C}^{(\mathrm p)}(\mathcal N)\leq E_{\mathrm C}(\mathcal N)$
for every $N_S$ \cite{Wilde2018}, and $E_{\mathrm C}(\sigma_{\mathcal N}(N_S))=E_{\mathrm F}(\sigma_{\mathcal N}(N_S))$
by Corollary~\ref{cor:superadd}, since $\sigma_{\mathcal N}(N_S)$ is a two-mode
Gaussian state.

\emph{Upper bound.} These channels can be simulated exactly by
continuous variable teleportation through a finite-energy two-mode Gaussian
resource state $\omega$, whose Gaussian entanglement of formation is
$\sup_{N_S}E_{\mathrm F}^{\mathrm G}(\sigma_{\mathcal N}(N_S))=\lim_{N_S\to\infty}E_{\mathrm F}^{\mathrm G}(\sigma_{\mathcal N}(N_S))$
\cite{TserkisDiasRalph2018}. Teleportation simulation gives
$E_{\mathrm C}(\mathcal N)\leq E_{\mathrm C}(\omega)\leq E_{\mathrm F}^{\mathrm G}(\omega)$ for the sequential cost
\cite{Wilde2018}, and $E_{\mathrm F}^{\mathrm G}=E_{\mathrm F}$ on two-mode Gaussian states by
Eq.~\eqref{eq:main} of the Letter. Thus, explicitly,
\begin{align}
 \sup_{N_S\geq0}E_{\mathrm C}(\sigma_{\mathcal N}(N_S))
 &\leq E_{\mathrm C}^{(\mathrm p)}(\mathcal N)
 \leq E_{\mathrm C}(\mathcal N)\nonumber\\
 &\leq\sup_{N_S\geq0}E_{\mathrm F}^{\mathrm G}(\sigma_{\mathcal N}(N_S)).
 \label{S:channelsandwich}
\end{align}
Corollary~\ref{cor:superadd} and Eq.~\eqref{eq:main} of the Letter identify the two endpoints for every finite $N_S$ before taking the supremum. This proves equality of the parallel and sequential costs with the supremum in Eq.~\eqref{S:channelcost}, without assigning an entanglement measure to an infinitely squeezed state.

\emph{Evaluation.} For a non-entanglement-breaking channel, the covariance matrix of $\sigma_{\mathcal N}(N_S)$ has the
squeezed thermal form of Sec.~\ref{S:specialsec} with
$a=N_S+\frac12$, $b=\tau(N_S+\frac12)+y$, $k=\sqrt{\tau N_S(N_S+1)}$, where
$y=(1-\eta)(N_B+\frac12)$, $(G-1)(N_B+\frac12)$ or $\xi$, and $\tau=\eta$, $G$ or
$1$, respectively. Dividing the defining equation of $r_0$ in
Eq.~\eqref{S:thermalformula} by $N_S$ and letting $N_S\to\infty$ gives
$(1+\tau)\cosh2r-2\sqrt\tau\sinh2r=2y$, i.e.,
\begin{equation}
 (1-\sqrt\tau)^2e^{2r}+(1+\sqrt\tau)^2e^{-2r}=4y .
\end{equation}
Its smaller root has
$e^{-2r}=[2y+\sqrt{4y^2-(1-\tau)^2}]/(1+\sqrt\tau)^2$, which equals
$\delta_{\mathcal N}$ in Eq.~\eqref{S:deltaN}, and $\delta_{\mathcal N}=\xi$ for $\tau=1$.
Since $\sinh^2r=(1-e^{-2r})^2/(4e^{-2r})$, the limit of
$E_{\mathrm F}(\sigma_{\mathcal N}(N_S))=g(\sinh^2r_0)$ is the last expression in
Eq.~\eqref{S:channelcost}. The condition $\delta_{\mathcal N}<1$ is equivalent to
$(1-\eta)N_B<\eta$, $(G-1)N_B<1$ and $\xi<1$, respectively, i.e., to $\mathcal N$
not being entanglement breaking. Entanglement-breaking channels admit measure-and-prepare implementations, hence have zero cost under LOCC. For the identity channel, $E_{\mathrm F}(\psi_{N_S})=g(N_S)\to\infty$, so the lower bound gives infinite unrestricted cost. A replacement channel has zero cost.
\end{proof}

For $N_B=0$, Eq.~\eqref{S:channelcost} gives
$E_{\mathrm C}(\mathcal L_\eta)=g(\eta/(1-\eta))$ and
$E_{\mathrm C}(\mathcal A_G)=g(1/(G-1))$, the values obtained in
Ref.~\cite{Wilde2018}, where the case $N_B>0$ was left open precisely because
$E_{\mathrm C}=E_{\mathrm F}$ was not known for the states $\sigma_{\mathcal N}(N_S)$. The equality of parallel and sequential costs means that allowing adaptive sequential tests does not increase the required entanglement rate for these channels.

A concrete irreversibility consequence follows analytically for weak additive noise. In nats per use, the two-way assisted entanglement-distribution capacity of $\mathcal T_\xi$ is at most $\xi-1-\ln\xi$ for $0<\xi<1$ \cite{PLOB2017}, while Eq.~\eqref{S:channelcost} gives
\begin{equation}
 E_{\mathrm C}(\mathcal T_\xi)-[\xi-1-\ln\xi]
 \longrightarrow 2-\ln4>0\qquad(\xi\to0).
\end{equation}
Indeed, $g(n)=\ln n+1+o(1)$ as $n\to\infty$, and $n=(1-\xi)^2/(4\xi)$. Consequently, for all sufficiently small positive $\xi$, simulating the channel consumes strictly more entanglement per use than can be distributed through it, even with two-way classical communication.

\section{Non-Gaussian states}
\label{S:nongaussian}

\subsection{Lower bound at fixed covariance matrix}
Let $\rho$ be any finite-energy two-mode state with covariance matrix $V$.
We prove Eq.~\eqref{eq:covbound} of the Letter,
\begin{equation}
 E_{\mathrm F}(\rho)\geq E_{\mathrm C}(\rho)\geq E_{\mathrm F}^{\mathrm G}(V).
 \label{S:covlower}
\end{equation}
The first inequality holds for every state, since
$E_{\mathrm C}(\rho)=\lim_m\frac1mE_{\mathrm F}(\rho^{\otimes m})$ and
$E_{\mathrm F}(\rho^{\otimes m})\leq mE_{\mathrm F}(\rho)$. For the second, if $V$ is
separable the right-hand side vanishes. Otherwise, apply local displacements
to center $\rho$, and the local Gaussian unitaries (and, if needed, the
exchange of parties) that bring $V$ to the canonical form $V_q+N_{t,u,v}$.
None of these operations changes $E_{\mathrm F}$ or $E_{\mathrm C}$. The expectation of
$\widehat D_t$ is a linear function of the second moments, so the transformed
$\rho$ has $\operatorname{Tr}(\rho\widehat D_t)=d_t(q)$, exactly as the Gaussian
state with the same covariance matrix. Corollary~\ref{cor:superadd}, applied
to $\rho^{\otimes m}$, whose $m$ pairs all have covariance matrix $V$, gives
$E_{\mathrm F}(\rho^{\otimes m})\geq mE_{\mathrm F}^{\mathrm G}(V)$. Since $\rho$ has finite
local entropy, $E_{\mathrm C}(\rho)$ is its regularized entanglement of formation
\cite{Hayden2001,Yamasaki2025}, and Eq.~\eqref{S:covlower} follows. (For the
bound on $E_{\mathrm F}$ alone, the case $m=1$, i.e., Theorem~1 averaged over any
pure-state decomposition of $\rho$, suffices; for $q=t<1$ one takes the
endpoint limit as in Sec.~\ref{S:thm1proof}.) Since the Gaussian state attains
the bound,
\begin{equation}
 \inf_{\rho:\,\mathrm{Cov}(\rho)=V}E_{\mathrm F}(\rho)
 =\inf_{\rho:\,\mathrm{Cov}(\rho)=V}E_{\mathrm C}(\rho)
 =E_{\mathrm F}^{\mathrm G}(V).
\end{equation}
In the same way, taking all $(q_j,t_j)$ equal in Eq.~\eqref{S:thm2} shows that
the quantitative witness \eqref{eq:mixedwitness} of the Letter, and hence the
balanced bound \eqref{eq:balancedbound}, are lower bounds on $E_{\mathrm C}(\rho)$
as well. A gap between $E_{\mathrm F}^{\mathrm G}(V)$ and the entanglement of
formation or cost of a non-Gaussian state can therefore only be closed using
information beyond second moments, or by physical preprocessing such as
local filtering (Sec.~\ref{S:filtersec}).

\subsection{Lossy photon-pair state with phase noise}
We generalize the example of Fig.~1(a) of the Letter by including local
phase noise. Independent pure-loss channels with transmissivities
$\eta_A,\eta_B$ act on $\sqrt{1-p}|00\rangle+\sqrt p|11\rangle$, followed by
local phase noise reducing the $|00\rangle\langle11|$ coherence by a
visibility $0\leq\kappa\leq1$. In the basis
$\{|00\rangle,|01\rangle,|10\rangle,|11\rangle\}$,
\begin{equation}
 \rho=\begin{pmatrix}
 r_{00}&0&0&z\\
 0&r_{01}&0&0\\
 0&0&r_{10}&0\\
 z&0&0&r_{11}
 \end{pmatrix},
\end{equation}
with $r_{00}=1-p+p(1-\eta_A)(1-\eta_B)$,
$r_{01}=p(1-\eta_A)\eta_B$, $r_{10}=p\eta_A(1-\eta_B)$,
$r_{11}=p\eta_A\eta_B$, and $z=\kappa\sqrt{\eta_A\eta_Bp(1-p)}$.
The exact entanglement of formation follows from the concurrence
$\mathcal C=2\sqrt{\eta_A\eta_B}\,[\kappa\sqrt{p(1-p)}-p\sqrt{(1-\eta_A)(1-\eta_B)}]_+$
\cite{Wootters1998}, so the state is entangled iff $\eta_A\eta_B>0$ and
$\kappa\sqrt{p(1-p)}>p\sqrt{(1-\eta_A)(1-\eta_B)}$.
The covariance matrix is of squeezed thermal form with
$a=\frac12+p\eta_A$, $b=\frac12+p\eta_B$, $k=z$, so the bound
\eqref{S:covlower} is given explicitly by Eq.~\eqref{S:thermalformula}, and
\begin{equation}
 \operatorname{Tr}(\rho\widehat D_t)=1+p\eta_A t^2+p\eta_B-2tz.
 \label{S:dephasedmoment}
\end{equation}

The bound is positive iff $\operatorname{Tr}(\rho\widehat D_t)<1$ for some
gain $t$ and mode ordering. Necessity follows by evaluating the supporting
inequality on the vacuum, for which $\langle\widehat D_t\rangle=1$;
sufficiency follows by letting $q\to0$ in Eq.~\eqref{eq:mixedwitness} of the
Letter, where the term $\lambda_{q,t}(1-\operatorname{Tr}\rho\widehat D_t)$ is
of order $q|\ln q|$ and dominates $\mathcal E(q)=O(q^2|\ln q|)$. For $p>0$, $z>0$ and $\eta_A\geq\eta_B>0$, minimizing over the allowed gains $0<t\leq1$ gives
\begin{equation}
 t_*=\min\!\left\{1,\frac{z}{p\eta_A}\right\}.
\end{equation}
The corresponding minimum is
\begin{equation}
 \min_{0<t\leq1}\operatorname{Tr}(\rho\widehat D_t)=
 \begin{cases}
 1+\eta_B[p-\kappa^2(1-p)],&z\leq p\eta_A,\\
 1+p(\eta_A+\eta_B)-2z,&z>p\eta_A.
 \end{cases}
\end{equation}
In the second branch, $z>p\eta_A\geq p\eta_B$ already makes the minimum smaller than one. In the first branch, this occurs exactly when $z^2>p^2\eta_A\eta_B$. Combining the branches, the covariance bound is positive iff $0<p<\kappa^2/(1+\kappa^2)$,
whereas the balanced measurement $t=1$ requires
$p<4\kappa^2\eta_A\eta_B/[(\eta_A+\eta_B)^2+4\kappa^2\eta_A\eta_B]$, a strictly
smaller region whenever $\eta_A\neq\eta_B$. For the parameters of
Fig.~1(a) of the Letter ($\kappa=1$, $\eta_A=0.9$, $\eta_B=0.2$) and
$p=0.01$, the optimized bound is $0.01759$ ebits, the balanced bound $0.01581$
ebits, and the exact value $0.01795$ ebits. By Eq.~\eqref{S:covlower}, both
bounds also hold for the entanglement cost.

\subsection{Photon-subtracted squeezed vacuum}
The state of Fig.~1(b) of the Letter is
\begin{equation}
 |\Psi_q^{-}\rangle=\frac{ab|\Psi_q\rangle}{\|ab|\Psi_q\rangle\|}
 =\sqrt{\frac{(1-q^2)^3}{1+q^2}}\sum_{n=0}^\infty(n+1)q^n|n,n\rangle,
\end{equation}
which is pure and non-Gaussian, with exact entanglement
$E(\Psi_q^-)=-\sum_n\pi_n\ln\pi_n$, where $\pi_n=(1-q^2)^3(n+1)^2q^{2n}/(1+q^2)$.
Its covariance matrix is symmetric and of squeezed thermal form, with
$a=b=\frac12+\bar n$ and $k=\langle ab\rangle$, where
\begin{equation}
 \bar n=\frac{2q^2(2+q^2)}{1-q^4}\,,\qquad
 \langle ab\rangle=\frac{2q(1+2q^2)}{1-q^4}.
\end{equation}
The balanced EPR expectation is therefore
\begin{equation}
 \delta_-=\langle\widehat D_1\rangle_{\Psi_q^-}
 =\frac{(1-q)(1-2q+3q^2)}{(1+q)(1+q^2)}
 =\frac{1-2q+3q^2}{1+q^2}\,\delta_0,
\end{equation}
where $\delta_0=(1-q)/(1+q)$ is the value for $|\Psi_q\rangle$. Since
$1-2q+3q^2<1+q^2$ for $0<q<1$, photon subtraction lowers the EPR
expectation at every squeezing level, in agreement with the
observations of Ref.~\cite{Kurochkin2014}. By Eq.~\eqref{eq:balancedbound}
of the Letter, which coincides with Eq.~\eqref{S:covlower} for this
symmetric covariance matrix, the certified entanglement of formation of
the subtracted state exceeds the entanglement $\mathcal E(q)$ of the
original Gaussian input. The bound captures about $80\%$ of the exact value
at $6$~dB of input squeezing, $93\%$ at $3$~dB, where it certifies $1.22$
ebits against $0.56$ ebits of the input, and approaches $100\%$ as the input
squeezing tends to zero. Symmetric losses of transmissivity $\eta$ applied after
subtraction map $\delta\mapsto1-\eta(1-\delta)$ for both states; since the
bound is monotone in $\delta$ and exact for the lossy Gaussian input, the
certified lower bound on $E_{\mathrm F}$ and $E_{\mathrm C}$ of the lossy subtracted
state still exceeds the common value $E_{\mathrm F}=E_{\mathrm C}$ of the equally lossy
Gaussian input, for $0<q<1$ and $0<\eta\leq1$.

\subsection{Local filtering}
\label{S:filtersec}
Local operations can modify the covariance matrix and reveal
entanglement invisible to second moments. Consider a local filter that
succeeds with probability $P_s$ and outputs a state $\rho_s$ with covariance
matrix $V_s$. Since the entanglement of formation does not increase on
average under local operations and classical communication
\cite{Bennett1996}, $E_{\mathrm F}(\rho)\geq P_sE_{\mathrm F}(\rho_s)$. The same holds
for the entanglement cost: applying the filter to each of $n$ copies of
$\rho$ and keeping the successful ones produces, with probability tending to
one, at least $(P_s-\epsilon)n$ copies of $\rho_s$ by LOCC, so any dilution
protocol for $\rho$ at rate $R$ yields one for $\rho_s$ at rate
$R/(P_s-\epsilon)$, whence $E_{\mathrm C}(\rho)\geq P_sE_{\mathrm C}(\rho_s)$. By
Eq.~\eqref{S:covlower} applied to the filtered state,
\begin{equation}
 E_{\mathrm F}(\rho)\geq E_{\mathrm C}(\rho)\geq P_s\,E_{\mathrm F}^{\mathrm G}(V_s).
 \label{S:weightedfilter}
\end{equation}
The factor $P_s$ is essential: conditioning alone could otherwise inflate
the certified value. As an example, consider the state of the lossy
photon-pair subsection and ideal noiseless attenuation
$K_j=\sigma_j^{n_j/2}$ on mode $j=A,B$, with $0<\sigma_j\leq1$, implementable
with a beam splitter and a heralded vacuum projection
\cite{Micuda2012,Hertz2021}. It maps
$(r_{01},r_{10},r_{11},z)\mapsto(\sigma_Br_{01},\sigma_Ar_{10},\sigma_A\sigma_Br_{11},\sqrt{\sigma_A\sigma_B}\,z)/P_s$,
with $P_s=r_{00}+\sigma_Br_{01}+\sigma_Ar_{10}+\sigma_A\sigma_Br_{11}$, and the
filtered covariance bound is positive iff
$z^2>(r_{10}+\sigma_Br_{11})(r_{01}+\sigma_Ar_{11})$. If the input is entangled,
$\Delta=z^2-r_{10}r_{01}>0$, and the choice
$\sigma_A=\sigma_B=\min\{1,\Delta/[2r_{11}(r_{10}+r_{01}+r_{11})]\}$ satisfies this
condition. With ideal filtering, the bound \eqref{S:weightedfilter}
therefore detects every entangled state of this family. Quantitatively,
at $p=0.45$, $\kappa=1$, $\eta_A=0.9$, $\eta_B=0.2$, the choice
$\sigma_A=0.2906$, $\sigma_B=1$ raises the certified value from $0.0054$ to
$0.0403$ ebits ($P_s=0.713$), against an exact value of $0.1703$ ebits.
For nonideal detectors, the calibrated conditional covariance matrix and
success probability are to be used in Eq.~\eqref{S:weightedfilter}, which
holds for any physical local instrument.

\subsection{Statistical and calibration aspects}
To measure $\widehat D_t$, the local phase references are chosen so that
$\langle ab\rangle$ is real and nonnegative, and first moments are subtracted
from the records. For a fixed gain, replacing the measured
$\langle\widehat D_t\rangle$ by a one-sided upper confidence limit yields a
conservative lower confidence bound on $E_{\mathrm F}$; scanning over gains,
or evaluating Eq.~\eqref{S:covlower} on a reconstructed covariance matrix,
requires a simultaneous confidence region. The certification is
device-dependent, since it assumes calibrated quadratures and a trusted
assignment of the modes, and is complementary to device-independent and
measurement-device-independent approaches \cite{Friis2019,Wang2025}. Finally,
since the logarithmic negativity and Gaussian convex-roof measures are
inequivalent orderings of entangled states \cite{AdessoIlluminati2005}, the
certified value of $E_{\mathrm F}$ cannot in general be inferred from
partial-transpose spectra.

\section{Entropy and Wigner--Yanase skew information}

Theorem~1 also constrains single-mode states, through their canonical
purification. For a single-mode state $\rho$ with finite mean energy, let
\begin{equation}
 \delta(\rho)=\|[a,\sqrt\rho]\|_2^2
 =I_{\rm WY}(\rho,x)+I_{\rm WY}(\rho,p),
 \label{S:deltaWY}
\end{equation}
where $I_{\rm WY}(\rho,X)=-\frac12\operatorname{Tr}([\sqrt\rho,X]^2)$ is the
Wigner--Yanase skew information \cite{WignerYanase1963}. The second equality
in Eq.~\eqref{S:deltaWY} follows from $a=(x+ip)/\sqrt2$, since the cross term
$\operatorname{Tr}([x,\sqrt\rho][p,\sqrt\rho])$ is real.

\begin{corollary}\label{cor:skew}
For every single-mode state with finite mean energy and $0<\delta(\rho)<1$,
\begin{equation}
 S(\rho)\geq g\!\left(\frac{(1-\delta)^2}{4\delta}\right),
 \label{S:skew}
\end{equation}
with equality for all displaced thermal states; for $\delta\geq1$ the bound
is replaced by zero.
\end{corollary}

\begin{proof}
Vectorize $C=\sqrt\rho$ as the normalized pure two-mode state
$|C\rangle\!\rangle$, whose reduced state is $\rho$. Its balanced EPR
expectation is $\|a^\dagger C-Ca^\dagger\|_2^2=\|[a,\sqrt\rho]\|_2^2=\delta$
(for $t=1$, $\langle\widehat D_1\rangle=G_1$), and Eq.~\eqref{eq:balancedbound}
of the Letter applies. For $\rho=(1-q^2)q^{2a^\dagger a}$, the purification
is $|\Psi_q\rangle$ and $\delta=(1-q)/(1+q)$, giving equality. Both $\delta$
and $S$ are invariant under displacements, which extends equality to all
displaced thermal states.
\end{proof}

The skew information $I_{\rm WY}(\rho,X)$ is the quantum part of the variance
of $X$: it equals the variance for pure states and decreases under mixing,
the remainder being classical uncertainty \cite{Luo2005}. Equation
\eqref{S:skew} can be inverted: if $\bar n_S$ denotes the mean photon number
of the thermal state with entropy $S(\rho)$, then
\begin{equation}
 \delta(\rho)\geq\Bigl(\sqrt{\bar n_S+1}-\sqrt{\bar n_S}\Bigr)^2 .
 \label{S:skewinv}
\end{equation}
A state can thus have small quantum quadrature fluctuations only by being
highly mixed, and at given entropy thermal states minimize them. For pure
states, $\delta=\operatorname{Var}(x)+\operatorname{Var}(p)\geq1$, consistently
with the bound being zero there. Squeezing a thermal state multiplies
$\delta$ by $\cosh2r$ at fixed entropy, so among Gaussian states equality
holds only for displaced thermal states. A related quantity, the quadrature
coherence scale, is built from commutators with $\rho$ rather than
$\sqrt\rho$ and has been used to quantify nonclassicality
\cite{DeBievre2019}.

The bound \eqref{S:skew} admits no counterpart from above: at any fixed
$\delta>0$ the entropy can be arbitrarily large. For instance, a
Fock-diagonal state whose photon-number distribution is a discretized
Gaussian of mean $M$ and variance $M/(4\delta_0)$ has
$\delta=\sum_n(n+1)(\sqrt{\lambda_{n+1}}-\sqrt{\lambda_n})^2\to\delta_0$ as
$M\to\infty$, while its entropy grows as $\frac12\ln M$. To keep $\delta$
exactly equal to $\delta_0$, note that the variances $M(1\pm\epsilon)/(4\delta_0)$,
with $0<\epsilon<1$, give $\delta\to\delta_0/(1\pm\epsilon)$, on opposite sides of
$\delta_0$; by continuity in the variance, an intermediate choice gives
$\delta=\delta_0$ for all large $M$, with entropy $\frac12\ln M+O(1)$. The region of the
$(\delta,S)$ plane accessible to single-mode states is therefore bounded only
from below, by the thermal curve. The asymmetric version of
Eq.~\eqref{S:skew} follows in the same way from Theorem~1 with $t<1$.

\section{Optimality of Gaussian quantum discord}
\label{S:discordsec}

\subsection{Classical correlations and discord}
The total correlations of a bipartite state $\rho_{AB}$ are quantified by the
mutual information $I(A{:}B)=S(\rho_A)+S(\rho_B)-S(\rho_{AB})$. Part of them
can be accessed by local measurements: if $B$ is measured with a POVM
$\{\Pi_k\}$, obtaining outcome $k$ with probability $p_k$ and leaving $A$ in
the conditional state $\rho_{A|k}$, the average reduction of the entropy of
$A$ is the information about $A$ gained by measuring $B$. Its maximum,
\begin{equation}
 J(A|B)=S(\rho_A)-\inf_{\{\Pi_k\}}\sum_kp_kS(\rho_{A|k}),
 \label{S:HV}
\end{equation}
is the classical correlation of Henderson and Vedral
\cite{HendersonVedral2001}. The quantum discord \cite{OllivierZurek2001}
\begin{equation}
 D(A|B)=I(A{:}B)-J(A|B)\geq0
 \label{S:discord}
\end{equation}
is the part of the correlations that no local measurement on $B$ can
access. It vanishes iff $\rho_{AB}=\sum_kp_k\,\rho_{A|k}\otimes|k\rangle\langle k|$
for some orthonormal basis $\{|k\rangle\}$ of $B$, i.e., iff the correlations
are classical from the point of view of $B$, and it can be nonzero for
separable states. For Gaussian states, restricting the infimum in
Eq.~\eqref{S:HV} to Gaussian measurements, i.e., general-dyne detections
with a pure Gaussian seed (including homodyne detection as a limit), defines
$J^{\rm G}$ and the Gaussian discord $D^{\rm G}=I-J^{\rm G}$, which admits a
closed formula for two-mode states \cite{AdessoDatta2010,GiordaParis2010}.
Clearly $J^{\rm G}\leq J$ and $D^{\rm G}\geq D$; whether Gaussian measurements
are optimal has been answered so far only for squeezed thermal and related
channel-generated states \cite{Pirandola2014}.

The link with the entanglement of formation is the Koashi--Winter monogamy
relation \cite{KoashiWinter2004}. If $\lvert\phi\rangle_{ABR}$ is a
purification of $\rho_{AB}$, every rank-one POVM on $B$ induces a pure-state decomposition of $\rho_{AR}$, and every pure-state decomposition arises in this way, allowing continuous outcomes. Refining a POVM into rank-one elements cannot increase the average conditional entropy, by concavity of entropy, so the optimization may be restricted to such measurements. Since the conditional state of $A$ is the reduced state of the corresponding pure component of $\rho_{AR}$, one finds
\begin{equation}
 J(A|B)=S(\rho_A)-E_{\mathrm F}(\rho_{AR}).
 \label{S:KW}
\end{equation}
The more $A$ is entangled with the purifying system $R$, the less
classical correlation it can share with $B$. This relation holds for
infinite-dimensional systems with $S(\rho_A)<\infty$ \cite{Shirokov2023}.

\subsection{Optimality of Gaussian measurements}
\begin{corollary}\label{cor:discord}
Let $\rho_{AB}$ be a two-mode Gaussian state whose
covariance matrix has a symplectic eigenvalue equal to $1/2$. Then
$J(A|B)=J^{\rm G}(A|B)$ and $D(A|B)=D^{\rm G}(A|B)$; the same holds with
$A$ and $B$ exchanged.
\end{corollary}

\begin{proof}
Let $V_{AB}=S(\tfrac12I\oplus\nu_{AB}I)S^{\mathsf T}$ be the
Williamson decomposition. Then $\rho_{AB}=U_S(|0\rangle\langle0|\otimes\tau_{\nu_{AB}})U_S^\dagger$,
and purifying the thermal state $\tau_{\nu_{AB}}$ with a TMSV yields a pure
three-mode Gaussian state $|\phi\rangle_{ABR}$ with $R$ a single mode, so
that $\rho_{AR}$ is a two-mode Gaussian state. By Eq.~\eqref{S:KW} and
Eq.~\eqref{eq:main} of the Letter,
$J(A|B)=S(\rho_A)-E_{\mathrm F}^{\mathrm G}(\rho_{AR})$.
It remains to show $J^{\rm G}(A|B)\geq S(\rho_A)-E_{\mathrm F}^{\mathrm G}(\rho_{AR})$,
i.e., that the optimal Gaussian decomposition of $\rho_{AR}$ is induced
by a Gaussian measurement on $B$. Since $\rho_{AR}$ is purified by the
single mode $B$, its covariance matrix also has the form
$V_{AR}=T(\tfrac12I\oplus\mu I)T^{\mathsf T}$, with $\mu=\sqrt{\det V_B}$. If $\mu=\frac12$, the mode $B$ is pure and uncorrelated with $AR$, so that $J(A|B)=J^{\rm G}(A|B)=0$; we therefore assume $\mu>\frac12$. Let $\gamma_p\leq V_{AR}$ be
a pure covariance matrix. In the Williamson frame,
$\gamma'=T^{-1}\gamma_pT^{-\mathsf T}\leq\tfrac12I\oplus\mu I$; its first
diagonal block $\gamma'_{11}$ satisfies $\gamma'_{11}\leq\frac12I$ and
$\det\gamma'_{11}\geq\frac14$, hence $\gamma'_{11}=\frac12I$, and positivity of
$(\tfrac12I\oplus\mu I)-\gamma'$ forces the off-diagonal blocks to vanish.
Thus $\gamma'=\frac12I\oplus\gamma_2$ with $\gamma_2$ a pure single-mode
covariance matrix and $\gamma_2\leq\mu I$. In the same frame,
$|\phi\rangle$ is the vacuum of the first Williamson mode times a pure
two-mode Gaussian state of the second Williamson mode and $B$, which up
to a local Gaussian unitary on $B$ is the TMSV with covariance
$\left(\begin{smallmatrix}\mu I&\zeta Z\\ \zeta Z&\mu I\end{smallmatrix}\right)$,
$\zeta^2=\mu^2-\frac14$, $Z=\mathrm{diag}(1,-1)$. General-dyne detection of
$B$ with pure seed $\varpi$ leaves the second mode in a pure Gaussian state
with covariance $\mu I-\zeta^2Z(\mu I+\varpi)^{-1}Z$, independent of the
outcome, while the first moments are Gaussian distributed
\cite{Weedbrook2012}. Given $\gamma_2$ with $\mu I-\gamma_2>0$, the choice
\begin{equation}
 \varpi=\zeta^2Z(\mu I-\gamma_2)^{-1}Z-\mu I
\end{equation}
reproduces it. A direct $2\times2$ computation using only
$\det\gamma_2=\frac14$ shows $\det\varpi=\frac14$ and
\begin{equation}
 \operatorname{Tr}\varpi=\frac{(\mu^2+\frac14)\operatorname{Tr}\gamma_2-\mu}{\det(\mu I-\gamma_2)}>0,
\end{equation}
since $\operatorname{Tr}\gamma_2\geq1$ makes the numerator at least $(\mu-\frac12)^2>0$.
Hence $\varpi$ is a valid pure seed; boundary cases with
$\det(\mu I-\gamma_2)=0$ are reached as homodyne limits. The induced
ensemble of $\rho_{AR}$ consists of displaced pure states with covariance
$\gamma_p$, each with entanglement $E(\gamma_p)$, which is the conditional
entropy of $A$. Choosing $\gamma_p$ optimal in Eq.~\eqref{eq:covopt} of the
Letter yields $J^{\rm G}(A|B)\geq S(\rho_A)-E_{\mathrm F}^{\mathrm G}(\rho_{AR})=J(A|B)$.
Together with $J^{\rm G}\leq J$, this proves the claim. The argument is
symmetric under $A\leftrightarrow B$.
\end{proof}

The class of states in Corollary~\ref{cor:discord} consists exactly of the two-mode reductions of pure
three-mode Gaussian states, equivalently of the states obtained by
applying to one arm of a pure two-mode Gaussian state a single-mode
Gaussian channel with a single-mode pure environment, followed by arbitrary
local Gaussian unitaries. It overlaps with, and is complementary to, the
family of Ref.~\cite{Pirandola2014}, which comprises squeezed thermal states
and related channel-generated states and was obtained from the
minimum output entropy of phase-insensitive channels. For a general
two-mode Gaussian state the purifying system $R$ consists of two modes,
and the same argument would require $E_{\mathrm F}=E_{\mathrm F}^{\mathrm G}$ for
$1\times2$-mode Gaussian states, which remains open. For the class in Corollary~\ref{cor:discord}, the many-copy statement follows from Theorem~2, as shown next.

\subsection{Regularized classical correlations and discord}
\begin{corollary}\label{cor:discordreg}
Under the hypothesis of Corollary~\ref{cor:discord},
\begin{equation}
 J(A^n|B^n)_{\rho_{AB}^{\otimes n}}=nJ^{\rm G}(A|B),\qquad
 D(A^n|B^n)_{\rho_{AB}^{\otimes n}}=nD^{\rm G}(A|B)
 \label{S:Jadd}
\end{equation}
for all $n\geq1$, and the same holds with $A$ and $B$ exchanged.
\end{corollary}

\begin{proof}
The purification $\lvert\phi\rangle_{ABR}^{\otimes n}$ of $\rho_{AB}^{\otimes n}$ has
$\rho_{AR}^{\otimes n}$ as its $AR$ marginal, a tensor power of a two-mode
Gaussian state. Eq.~\eqref{S:KW} and Corollary~\ref{cor:superadd} give
$J(A^n|B^n)=nS(\rho_A)-nE_{\mathrm F}(\rho_{AR})=nJ(A|B)$, and
Corollary~\ref{cor:discord} gives $J(A|B)=J^{\rm G}(A|B)$. The statement on
discord follows since the mutual information is additive.
\end{proof}

The regularized classical correlation is therefore attained by Gaussian
measurements on single copies: for these states, no collective measurement
on many copies extracts more classical correlation per copy.

\end{document}